\documentclass[11pt,a4paper]{article}
\pdfoutput=1
\usepackage{jheppub}
\usepackage{braket}
\usepackage{amsmath}
\usepackage{verbatim}
\usepackage{amssymb}
\usepackage{bbold}
\usepackage{inputenc, array}
\usepackage{textcomp}
\usepackage{appendix}
\usepackage[dvipsnames]{xcolor}
\usepackage{epsfig}
\usepackage{mathrsfs}

\newcommand{\xcentcolon}

\usepackage{color}
\newcommand\anote[1]{\textcolor{blue}{\bf [AB:\,#1]}}

\newcommand{\e}{\epsilon}
\newcommand{\be}[1]{\begin{equation}\label{#1} }
\newcommand{\ee}{\end{equation}}
\newcommand{\bea}[1]{\begin{eqnarray}\label{#1} }
\newcommand{\eea}{\end{eqnarray}}
\newcommand{\p}{\partial}
\newcommand{\refb}[1]{(\ref{#1})}

\newcommand{\eps}{\varepsilon}

\DeclareMathOperator{\extdm}{d}
\newcommand{\extd}{\extdm \!}

\renewcommand{\a}{\alpha}

\renewcommand{\b}{\beta}
\renewcommand{\t}{\tau}
\newcommand{\s}{\sigma}

\title{Boundary Carroll CFTs: SUSY and Superstrings}
\author[a]{Arjun Bagchi,} \author[b]{Shankhadeep Chakrabortty,}\author[b]{Pronoy Chakraborty,} \author[b,c]{Ritankar Chatterjee,} \author[a,d]{and Priyadarshini Pandit.}
\author{\\}

\affiliation[a]{Indian Institute of Technology Kanpur, Kanpur 208016, INDIA.\\} 

\affiliation[b]{Indian Institute of Technology Ropar, Punjab 140001, INDIA. \\}

\affiliation[c]{Beijing Institute of Mathematical Sciences and Applications, Beijing 101408, China. \\}

\affiliation[d]{Tata Institute of Fundamental Research, Homi Bhabha Rd, Mumbai 400005, India.\\}

\emailAdd{abagchi@iitk.ac.in, s.chakrabortty@iitrpr.ac.in, pronoy.22phz0009@iitrpr.ac.in, staff.ritankar.chatterjee@iitrpr.ac.in, priyadarshini.pandit@tifr.res.in}

\preprint{}

\abstract{We consider two dimensional superconformal Carrollian theories with boundaries and construct two variants of the Boundary Superconformal Carrollian Algebra (BSCCA), viz. the Homogeneous and the Inhomogeneous, by making appropriate identification of the parent superconformal Carrollian algebras. These new algebras are then recovered by appropriate limits of a single copy of Super Virasoro algebra. We then focus on the theory of null tensionless superstrings and construct, for the first time, an open null superstring. The Homogeneous version of the BSCCA is realised as worldsheet symmetries on this open null superstring.}

\begin{document}
\maketitle
\newpage
\section{Introduction}
Carrollian symmetries are ubiquitous. From the null boundary of asymptotically flat spacetimes (AFS) and hence the constructions of holography for AFS \cite{Bagchi:2010zz, Bagchi:2016bcd, Donnay:2022aba, Bagchi:2022emh, Donnay:2022wvx, Bagchi:2023fbj} to the horizons of black holes \cite{Donnay:2019jiz}, Carroll and Conformal Carroll algebras appear on generic null surfaces \cite{Duval:2014uva, Duval:2014lpa}. Carroll symmetry has found uses in condensed matter physics in the context of fractons \cite{Bidussi:2021nmp}, systems with flat bands \cite{Bagchi:2022eui} and at phase separation \cite{Biswas:2025dte}, in hydrodynamics in the context of highly boosted fluids \cite{Bagchi:2023ysc, Bagchi:2023rwd}, and also in the context of cosmology and dark energy \cite{deBoer:2021jej}. For a detailed account of Carrollian symmetries and its uses in various physical systems, the reader is pointed to the recent review \cite{Bagchi:2025vri}.

\medskip

In this work, we focus on Carrollian Conformal Field Theories (CCFTs) in two spacetime dimensions (2d). In recent work \cite{Bagchi:2024qsb}, 2d CCFTs with boundaries was explored and a new algebra called the Boundary Carroll Conformal Algebra (BCCA) was discovered. 

\medskip

Boundary CFTs have been extremely useful in the study of defects and impurities in condensed matter systems starting with the seminal work of Affleck and Ludwig in understanding the Kondo model \cite{Affleck:1990zd, Affleck:1990iv}. Other prominent examples are quantum wires \cite{vanWees:1988zz} and quantum dissipative systems \cite{Johanesson:2003}. In the context of quantum gravity and holography, BCFTs form the core of the AdS/BCFT correspondence \cite{Takayanagi:2011zk, Fujita:2011fp} where a brane is introduced in the bulk Anti de Sitter spacetime to cap off the spacetime and thus form a boundary on the dual CFT. BCFTs are also central to the understanding of open strings and D-branes in string theory. It is expected that Boundary CCFTs would play analogous roles in terms of understanding condensed matter systems and quantum gravity (e.g. in a AFS/BCCFT correspondence, where branes are introduced in AFS).

\medskip 

In the present work, we extend this recent construction of BCCFT to its supersymmetric version. Supersymmetric CCFTs have previously been explored in \cite{Barnich:2014cwa, Lodato:2016alv, Bagchi:2016yyf, Bagchi:2017cte, Bagchi:2018wsn, Bagchi:2022owq, Zheng:2025cuw, Zheng:2025rfe}. In the first part of our paper, we will construct a Boundary Superconformal Carrollian field theory in $d=2$.


\medskip

One of the significant uses of 2d CCFTs is in the context of tensionless null strings where these symmetries arise on the null worldsheet \cite{Bagchi:2013bga}. Building on earlier work, and the identification of the worldsheet symmetries with these non-Lorentzian structures \cite{Schild:1976vq, Isberg:1993av}, a body of work has built up over the past years which studies null strings from the point of view of these Carrollian worldsheet symmetries \cite{Casali:2016atr,Bagchi:2020fpr, Bagchi:2021rfw, Bagchi:2019cay, Chen:2023esw, Banerjee:2023ekd, Banerjee:2024fbi, Chen:2025gaz}. One of the outstanding questions had been the construction of open strings in this context and in \cite{Bagchi:2024qsb}, this was answered and it was shown that in the Dirichlet version of the open null string, the BCCA arises as worldsheet symmetries. In the second part of our paper, we will consider null superstrings and construct, for the first time, an open null superstring. We will show how the symmetries on the worldsheet of this string become the algebra that we discover in the first part of the paper, viz the Boundary Superconformal Carroll Algebra. 

\medskip

Our findings in this work are presented in two main parts: The first part of the paper is devoted to the construction of the Boundary Superconformal   Carrollian Algebra (BSCCA) from a field-theoretic perspective. In Section \ref{BCCFT}, we review the basics of 2d Carrollian CFTs, obtained both from the limit of relativistic CFTs and from a geometric formulation. We then introduce boundaries in Carrollian manifolds and derive the associated algebra, called the Boundary Carrollian Conformal Algebra (BCCA). We also comment on the choice of boundary generators for the bosonic BCCA. Section \ref{CSCFT} discusses the supersymmetric extension of 2d Carrollian CFTs. We recall from earlier works that there are two distinct versions of superconformal Carrollian CFTs, namely the Homogeneous and Inhomogeneous versions, arising from different In\"on\"u–Wigner contractions of the supersymmetric Virasoro algebra. In Section \ref{BasantaBiswas}, we study the effect of boundaries in superconformal Carrollian theories and obtain the Boundary Superconformal Carrollian Algebras (BSCCAs), i.e. the Homogeneous and Inhomogeneous SCCAs. Furthermore, we show that both these boundary algebras can also be obtained from the Carrollian limit of a single copy of the super Virasoro algebra, expressed in different basis.

\medskip

The second part of the paper focuses on an application of the above construction, namely the appearance of the Homogeneous BSCCA as the residual symmetry algebra on the worldsheet of null open strings. Section \ref{bosonic null strings} revisits the construction of both closed and open bosonic null strings, worldsheets of which display 2d CCA and 2d BCCA respectively. Section \ref{null superstrings} develops the classical theory of null open superstrings in detail, supplemented by a brief review of tensile open superstrings in Appendix \ref{tensile superstring}. This section begins with a short review of closed null homogeneous superstrings. Then we proceed to discuss boundary conditions in open null strings. After that we take a closer look at the superspace formulation of the theory, apply boundaries in the superspace and find out the subset of BSCCA generators in the Neveu Schwarz sector which survives after applying the boundary. We construct the energy momentum tensor modes in terms of the bosonic and fermionic modes and show that the energy momentum tensor modes give us the same BSCCA symmetry algebra. Finally, we conclude this section by showing how this string emerges at the tensionless limit of tensile superstring. We perform this limiting analysis from both from mode expansion perspective as well as from superspace vector field representation perspective.

\medskip

In Section \ref{conclusions} we conclude with a summary of our results and a discussion on future directions. Appendix \ref{tensile superstring} reviews tensile open superstring theory in detail. Here we also cover a less emphasised aspect of tensile open superstring; namely the superspace formulation of open superstring after the introduction of boundaries in the superspace. Nevertheless, it is important to note that worldsheet theory of open tensile/tensionless superstrings corresponds to some specific boundaries in superspace. More generally several other boundaries can be considered to study boundary Superconformal (or boundary Carrollian Superconformal) field theories. Appendix \ref{different set if boundaries in superspace} explores all these different possibilities for both relativistic as well as for homogenous Carrollian superspace.

\newpage

\section{Boundary Carroll CFTs: A recap of the Bosonic story}\label{BCCFT}

In the first part of our paper, we will focus on 2d Carroll CFTs, their supersymmetric extensions and how one introduces boundaries in these Carroll SCFTs. After a quick review of basics of 2d CCFTs, we will remind the reader of the bosonic version of Boundary CCFTs, specifically the emergence of a new algebra the Boundary Carroll Conformal Algebra (BCCA). We then recap the supersymmetric extensions of CCFTs and the fact that there are two distinct types of Superconformal Carrollian Field Theories (SCCFTs) that emerge from different In\"on\"u-Wigner contractions of the parent 2d relativistic SCFTs. We will then go on to our main results in this part of the paper, where we introduce boundaries in SCCFTs, leading to new two classes of Boundary Supersymmetric Carrollian Conformal Algebras (BSCCA). 

\medskip

We begin in this section with a review of the bosonic version of conformal Carrollian symmetries, first the vanilla kind and then flavoured with boundaries. 

\subsection{2d Carrollian CFTs} 
We now briefly recap salient features of 2d CCFTs, viz. how to obtain it from a limit of relativistic 2d CFTs and a more geometric formulation of Conformal Carrollain symmetries in terms of conformal isometries of a Carrollian manifold. 

\subsection*{From Relativistic to Carroll CFTs}

The power of conformal symmetry manifests itself in all its glory in  relativistic conformal field theories in $d=2$.  It is well known that in 2d, there exists an infinite set of local conformal transformations generated by two copies of the Virasoro algebra
\begin{equation}
  \begin{split}
      [\mathcal{L}^\pm_n,\mathcal{L}^\pm_m]=&(n-m)\mathcal{L}^\pm_{n+m}+\frac{c^\pm}{12}n(n^2-1)\delta_{n+m},\quad [\mathcal{L}^+_n,\mathcal{L}^-_m]=0.
  \end{split}  
\end{equation} 
Here $c^\pm$ are the central charges of the Virasoro algebra. The infinite underlying symmetry of the relativistic 2d CFT helps is understanding these theories even without recourse to any underlying Lagrangian description. 

\medskip

We now move to Carrollian CFTs. Carroll symmetries appear in the vanishing speed of light limit $c$ of relativistic theories and an In\"on\"u-Wigner $c\to0$ contraction of the Poincar\'e symmetry  results in the Carroll algebra \cite{Leblond65, SenGupta:1966qer}. To understand how the same happens for a relativistic 2d CFT, let us focus on the generators of the Virasoro algebra defined on a cylinder parametrized by $(\t, \s)$:
\begin{align}
\mathcal{L}^\pm_n=ie^{in\s^\pm}\partial_\pm,
    \label{virvector}
\end{align}
where $\partial_{\pm}=\frac{1}{2}(\partial_\tau\pm\partial_\sigma)$. The Carroll contraction of these Virasoro generators amounts to taking the following limit
\begin{align}
 \tau\to \e\tau, \sigma \to \sigma~ \text{with}~ \e \to 0.
\end{align}
This is equivalent to taking the speed of light to zero and results in the emergence of Carrollian Conformal generators in 2d as
\begin{align}\label{cylinder}
    L_n &=i\,e^{in\sigma}(\partial_\sigma+in\tau\,\partial_\tau),\quad M_n= i\,e^{in\sigma}\partial_\tau,
\end{align}
where $\tau$ is the null direction $\mathbb{R}_{null}$ and $\sigma$ parametrizes the angular coordinate of $S^1$. In terms of the Virasoro generators, this amounts to the contraction 
\begin{equation}\label{Carrcontraction}
    L_n=\mathcal{L}^+_n-\mathcal{L}^-_{-n},\quad\quad M_n=\epsilon(\mathcal{L}^+_n+\mathcal{L}^-_{-n}).
\end{equation}
These generators \eqref{cylinder} satisfy the 2d Carrollian Conformal Algebra (CCA$_2$):
 \begin{subequations}\label{bms3}
  \begin{align}
    \left[L_n,L_m\right]&=(n-m)L_{n+m} +\frac{c_L}{12}(n^3-n)\delta_{n+m}, \\
    \left[L_n,M_m\right]&=(n-m){M}_{n+m}+\frac{c_M}{12}(n^3-n)\delta_{n+m},\\
    \left[M_n, M_m\right]&=0.
\end{align}  
\end{subequations}
Here, $c_L$ and $c_M$ are the central charges. This algebra isomorphic to $3$d Bondi-Metzner-Sachs algebra (BMS$_3$) algebra \cite{Barnich:2006av}, which is the asymptotic symmetry algebra of asymptotically flat 3d spacetimes at the null boundary \cite{Bondi:1962px, Sachs:1962wk}. This isomorphism between the conformal Carroll and the BMS algebras extend to arbitrary dimensions, forming the building blocks of holography in asymptotically flat spacetimes \cite{Bagchi:2010zz, Duval:2014uva, Duval:2014lpa}. 

\subsection*{The geometric formulation}

We now move on to discuss the geometric aspect of the Carrollian and Conformal Carrollian symmetries \cite{Duval:2014lpa, Duval:2014uva, Duval:2014uoa}. The Carrollian algebra corresponds to the isometry algebra of a flat Carrollian manifold. A Carrollian manifold in $d$-dimensional spacetime is defined by a degenerate symmetric rank-two covariant tensor field $h_{\mu\nu}$ and a nowhere-vanishing vector field 
$v^\mu$  that spans the 1-dimensional kernel of $h_{\mu\nu}$, i.e, $v^{\mu}h_{\mu\nu}=0$. The isometries of a Carroll manifold are generated by vector fields $\xi^\mu$ that preserve the following Carrollian structure
\begin{equation}
    \mathcal{L}_\xi v^\mu = 0,\qquad\qquad \mathcal{L}_\xi h_{\mu\nu} = 0.
\label{eq:angelinajolie}
\end{equation}
where $\mathcal{L}_\xi$ denotes the Lie derivative along $\xi^\mu$. When the Carroll structure is flat, then the Lie algebra of these vector fields close to form the Carroll algebra. Here in this work, we shall mostly be interested in the conformal structure of Carroll manifolds where the conformal isometries are generated by vector fields that satisfy the Carroll conformal Killing equations,
\begin{equation}
    \mathcal{L}_\xi v^\mu = \lambda\, v^\mu,\qquad\qquad \mathcal{L}_\xi h_{\mu\nu} = - 2 \lambda\, h_{\mu\nu}\,.
\label{eq:angelinajolie1}
\end{equation}
Here $\lambda$ is the conformal factor. Solution to the above conformal isometry equations \eqref{eq:angelinajolie1} in 2d for flat Carroll manifolds (with coordinates $\tau$ and $\sigma$ so that $v^\tau=1$, $h_{\sigma\sigma}=1$, $v^\sigma=h_{\tau\tau}=h_{\tau\sigma}=0$) yields the Carroll conformal Killing vectors given as
\begin{equation}   \xi=[\xi_M(\sigma)+\tau\xi_L^\prime(\sigma)]\partial_\tau+\xi_L(\sigma)\partial_\sigma.
\end{equation} 
Here, $\xi_M(\sigma)$ and $\xi_L(\sigma)$ are arbitrary functions of spatial coordinate $(\sigma)$ parameterizing ``supertranslations'' and ``superrotations'' respectively. The Lie brackets generate the (centerless) $d$-dimensional conformal Carroll algebra. The generators of the algebra of Carrollian conformal isometry can be defined as 
\begin{equation}
L(\xi_L)=\tau\,\xi_L^\prime(\sigma)\,\partial_\tau+\xi_L(\sigma)\,\partial_\sigma, \quad\quad M(\xi_M)=\xi_M(\sigma)\partial_\tau.
\end{equation}
Expanding these functions in terms of Fourier modes:
$\xi_L=\sum a_n\, e^{in\sigma},~~\xi_M=\sum b_n\,e^{in\sigma}$, the generators can be re-written as 
\begin{equation} \label{fouriermodes}
\begin{split}
L(\xi_L)&=\sum_n a_n e^{in\sigma}(\partial_\sigma+in\tau\,\partial_\tau)=-i\sum_na_n\,L_n,\\ M(\xi_M)&=\sum_n b_n e^{in\sigma}\partial_\tau=-i\sum_n b_n\,M_n.
\end{split}
\end{equation}
It is evident from \eqref{fouriermodes} form of the generators are same as \eqref{cylinder} and permitting central extensions obey the 2d Carrollian Conformal algebra or BMS$_3$ algebra \eqref{bms3}. The generators can be understood in terms of the geometry of 3d flat space on its null boundary $\mathscr{I}^\pm$ with cylinder topology $\mathbb{R}_{null}\times S^1$, where the $S^1$ is sometimes referred to as ``celestial circle''. The superrotations $L_n$ generate diffeomorphisms of this celestial circle. The supertranslations $M_n$ generate angle-dependent translations along the null direction, i.e.~along advanced or retarded time.  

\subsection{2d Carrollian CFTs with boundaries}\label{CCA} 
\begin{figure}[htbp]
\centering
    \includegraphics[scale=0.65]{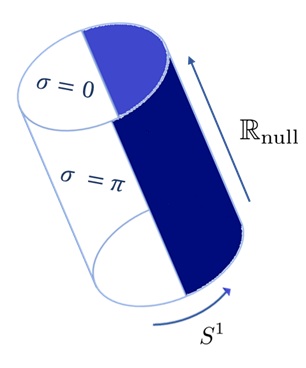}
    \caption{Null cylinder with boundaries}
    \label{fig2}
\end{figure}

We now consider 2d CCFTs with boundaries. Specifically, we study these theories defined on the null cylinder with Carrollian structure
\begin{equation}
    v^\mu\partial_\mu=\partial_\tau\quad\qquad \extd s^2 = 0\cdot \extd\tau^2 + \extd\sigma^2\quad(\sigma\sim\sigma+2\pi)\,.
\end{equation}
We remind the reader that the generators of 2d CCFT on the cylinder are given by \eqref{cylinder}. We now introduce boundary on the cylinder at $\sigma=0$ and $\sigma=\pi$, as illustrated in Fig~\ref{fig2}.
 \medskip

\subsection*{Boundary Carroll Conformal Algebra}

From the figure, it can be clearly seen that the translation symmetry along the $\sigma$ direction is broken. Hence the operators with a non-zero $\partial_\sigma$ component at the boundaries cannot preserve our boundary conditions and have to be excluded from the algebra. Since $L_n$'s in \eqref{cylinder} contain terms proportional to $\partial_\sigma$, we need to construct a new set of generators whose $\partial_\sigma$ terms vanish appropriately at $\sigma=0,\pi$. Hence we define
\begin{equation}
\label{bccft21}
    \mathcal{O}_n = L_n-L_{-n} = 2i\sin(n\sigma)\,\partial_\sigma+2in\tau\cos(n\sigma)\,\partial_\tau\,.
\end{equation}
At the boundaries $\sigma=0$ and $\pi$, this reduces to 
\begin{equation}\label{bvectorfieldrep1}
   \mathcal{O}_n=\pm 2in\tau\,\partial_\tau~~ \text{for}~~ \sigma=0 ~\text{and}~\pi, 
\end{equation}
thus preserving the boundary conditions. Similarly, we redefine
\begin{equation}
\mathcal{P}_n = M_n + M_{-n} = 2 \cos(n\sigma)\, \partial_\tau,
\label{eq:whatever}
\end{equation} 
where $\mathcal{O}_{-n}=-\mathcal{O}_n$ and $\mathcal{P}_{-n}=\mathcal{P}_n$, $\forall{n}\in\mathbb{Z}$. At this juncture, there could have been different choices of the supertranslation generator we could have chosen, e.g. 
\begin{align}
 \mathcal{R}_n = M_n - M_{-n} = 2 \sin(n\sigma)\, \partial_\tau,   
\end{align}
or chosen to keep both halves and hence the original $M$. In our explicit construction of the open null string later in the paper, it is $\mathcal{P}$ that would show up and hence we will focus on this. But we comment on the other choices later in the section in more detail. 

\medskip

We now compute the symmetry algebra satisfied by these boundary Carrollian CFT generators that preserve the boundary conditions we imposed on the Carrollian null cylinder. The resulting symmetry algebra called the Boundary Conformal Carrollian Algebra (BCCA) \cite{Bagchi:2024qsb} is given by 
\begin{subequations}\label{bdyCarr}
  \begin{align}
    \left[\mathcal{O}_n,\,\mathcal{O}_m\right]&=(n-m)\mathcal{O}_{n+m}-(n+m)\mathcal{O}_{n-m},\\
    \left[\mathcal{O}_n,\,\mathcal{P}_m\right]&=(n-m)P_{n+m}+(n+m)P_{n-m}+\frac{c_M}{12}(n^3-n)(\delta_{n+m}+\delta_{n-m}),\\
    \left[\mathcal{P}_n,\,\mathcal{P}_m\right]&=0.
\end{align}  
\end{subequations}
Here we see that there is only one possible central extension $c_M$ in the cross commutator. It is interesting to note that we can use a single set of generators for the entire algebra at the cost of being somewhat more cumbersome. This new basis is defined by 
\begin{align}
Q_n=\mathcal{O}_n+\mathcal{P}_n
\end{align}
The algebra now reads: 
\begin{align}
   [Q_n,Q_m]&=(n-m)\left(\frac32Q_{n+m}+\frac12Q_{-n-m}\right)+(n+m)\left(\frac12Q_{m-n}-\frac12Q_{n-m}\right) \nonumber \\
   & \quad +\frac{c_P}{6}(n^3-n)\delta_{n+m,0}.  
\end{align}
When we refer to the BCCA, we will however focus on the previous version, viz. \eqref{bdyCarr}. 

\subsection*{BCCA from a single Virasoro}

We now show how to obtain the BCCA as a novel contraction of a single copy of the Virasoro algebra. To this end, we  express the Virasoro generators in a different basis (useful when boundaries are introduced, as we will see below) defined as
\begin{equation}
    \mathbb{L}_n=\mathcal{L}^+_n+\mathcal{L}^-_n,\qquad\qquad \widetilde{\mathbb{L}}_n=\mathcal{L}^+_n-\mathcal{L}^-_n.
\end{equation}
In this basis, the Virasoro algebra take the following form :
\begin{subequations}\label{pro97}
\begin{align}
    [\mathbb{L}_{m},\mathbb{L}_{n}]&=(m-n)\mathbb{L}_{(m+n)}+\frac{\mathbb{c}}{12} m(m^2-1)\delta_{m+n,0}\\
    [\widetilde{\mathbb{L}}_{m},\widetilde{\mathbb{L}}_{n}]&=(m-n)\mathbb{L}_{(m+n)}+\frac{\mathbb{c}}{12} m(m^2-1)\delta_{m+n,0}\\
     [\mathbb{L}_{m},\widetilde{\mathbb{L}}_{n}]&=(m-n)\widetilde{\mathbb{L}}_{m+n}+\frac{\tilde{\mathbb{c}}}{12} m(m^2-1)\delta_{m+n,0}.
\end{align}
\end{subequations}
where $\mathbb{c}, \tilde{\mathbb{c}}=c^+\pm c^-$. Using \eqref{virvector}, the coordinate representation for these modified generators on the cylinder are 
\begin{subequations}
\begin{align}\label{pro911}
    \mathbb{L}_{n}=-\frac{1}{2}\bigl\{e^{i n(\tau+\sigma)}(\partial_{\tau}+\partial_{\sigma})+ e^{i n(\tau-\sigma)}(\partial_{\tau}-\partial_{\sigma})\bigr\}\\
    \widetilde{\mathbb{L}}_{n}=-\frac{1}{2}\bigl\{e^{i n(\tau+\sigma)}(\partial_{\tau}+\partial_{\sigma})- e^{i n(\tau-\sigma)}(\partial_{\tau}-\partial_{\sigma})\bigr\}
\end{align}
\end{subequations}
The generators $\mathbb{L}_{n}$, $\widetilde{\mathbb{L}}_{n}$ as given in \eqref{pro911} at $\s=0, \pi$, respectively become the following
\begin{align}\label{sigmazero}
    \mathbb{L}_{n}|_{\s=0}=-e^{in\tau}\partial_{\tau},\,~  
    \widetilde{\mathbb{L}}_{n}|_{\s=0}=e^{in\tau}\partial_\s; \qquad 
    \mathbb{L}_{n}|_{\s=\pi}=e^{in\tau}\partial_{\tau},\,~ 
    \widetilde{\mathbb{L}}_{n}|_{\s=\pi}=-e^{in\tau}\partial_\s.
\end{align}
We thus notice that $\mathbb{L}_{n}$s respect both $\s=0, \pi$ boundaries  while $\widetilde{\mathbb{L}}_{n}$s are not compatible with them (since these contain $\partial_\s$). Looking at \eqref{pro97}, one can see that boundary compatible symmetry generators form one copy of the Virasoro algebra. Now we can reproduce the boundary Carrollian conformal generators previously introduced in \eqref{bccft21}, and in \eqref{eq:whatever} by taking the following ultra-relativistic contraction of the Virasoro generators in the new basis,
\begin{equation}
    \mathcal{O}_n = \mathbb{L}_n - \mathbb{L}_{-n},\qquad \mathcal{P}_n = \epsilon \,(\mathbb{L}_n + \mathbb{L}_{-n}),\qquad \epsilon \to 0.
    \label{eq:contraction}
\end{equation}
This of course is true at the level of the algebra as well, without recourse to the explicit form of the generators on the null cylinder. It can be easily checked that the above contraction reduces a single copy of the Virasoro algebra, here given by generators $\mathbb{L}_n$, to the BCCA.

\subsection*{Choice of boundary generators for BCCA}

We now comment on the choice of the  generators for the bosonic boundary CCA. We have seen above that the choice of boundary generators was 
\begin{align}
    \mathcal{O}_n = L_n - L_{-n}, \quad \mathcal{P}_n = M_n + M_{-n}
\end{align}
The choice of the combinations of $L$s was dictated by the fact that we wanted to put boundaries on constant $\sigma$, viz. at $\sigma=0, \pi$ on the null cylinder where the field theory was defined. A priori, the choice of combination of $M$ seemed arbitrary and we promised a more detailed account of the other choices, viz 
\begin{align}
    \mathcal{R}_n = M_n - M_{-n}
\end{align}
and unrestricted $M$s. However, using limiting analysis from the Virasoro generators ($\mathbb{L},\widetilde{\mathbb{L}}$), we can clearly see that two of the above three ($\mathcal{R}$ and $M$) are not allowed. 
\begin{subequations}
\begin{align}
   \mathcal{P}_n=M_n+M_{-n}&=\epsilon\{(\mathcal{L}^+_n+\mathcal{L}^-_{-n})+(\mathcal{L}^+_{-n}+\mathcal{L}^-_{n})\} =\epsilon\, (\mathbb{L}_n+\mathbb{L}_{-n})\\   
   \mathcal{R}_n=M_n-M_{-n}&=\epsilon\{(\mathcal{L}^+_n+\mathcal{L}^-_{-n})-(\mathcal{L}^+_{-n}+\mathcal{L}^-_{n})\}=\epsilon\, (\widetilde{\mathbb{L}}_n-\widetilde{\mathbb{L}}_{-n})\\
    M_n&=\epsilon\,(\mathcal{L}^+_n+\mathcal{L}^-_{-n}) =\frac{\epsilon}{2}(\mathbb{L}_n+\mathbb{L}_{-n}+\widetilde{\mathbb{L}}_n-\widetilde{\mathbb{L}}_{-n}).
   \end{align}
\end{subequations}
As we already know from the expression of $\mathbb{L}, \widetilde{\mathbb{L}}$ in \eqref{sigmazero} at $\sigma=0,\pi$, due to the presence of $\partial_\sigma$ term, $\widetilde{\mathbb{L}}$ is not allowed at the boundaries. Hence, any generators containing $\widetilde{\mathbb{L}}$ as in $\mathcal{R}, M$, are not allowed.

\medskip

It is interesting to consider what would happen if we had disregarded this limiting construction. Although, it may seem at first glance \cite{Bagchi:2024qsb}, that the algebras with $\mathcal{P}$ and $\mathcal{R}$ are identical, there are crucial differences. This stems from the critical behaviour: 
\begin{align}
    \mathcal{P}_n=\mathcal{P}_{-n}, \quad \mathcal{R}_n=-\mathcal{R}_{-n}.
\end{align}
So, although the non-centrally extended infinite dimensional algebra stays the same, the central terms differ
\begin{subequations}
    \begin{align}
        &[\mathcal{O}_n,\mathcal{P}_m]=(n-m)\mathcal{P}_{n+m}+(n+m)\mathcal{P}_{n-m} + \frac{c_m}{12} (n^3-n)(\delta_{n+m}+\delta_{n-m}) \\&[\mathcal{O}_n,\mathcal{R}_m]=(n-m)\mathcal{R}_{n+m}+(n+m)\mathcal{R}_{m-n}+ \frac{c_r}{12} (n^3-n)(\delta_{n+m}+\delta_{n-m})
    \end{align}
\end{subequations}
In particular, 
\begin{align}
    c_m = c_M, \quad c_r = 0. 
\end{align}
The second equality follows in a straightforward manner by noticing
\begin{align}
        [\mathcal{O}_n,\mathcal{R}_n]=\frac{c_r}{12}(n^3-n), \, [\mathcal{O}_n,\mathcal{R}_{-n}]=\frac{c_r}{12}(n^3-n), \, [\mathcal{O}_n,\mathcal{R}_n]=- [\mathcal{O}_n,\mathcal{R}_{-n}] \implies c_r=0.
\end{align}
So the BCCA with $\mathcal{P}$ allows for non-trivial central extensions, while the one with $\mathcal{R}$ does not. 

\medskip

We close this section with comments on the finite dimensional algebra generated by the two different generators. The global part of 2d Conformal Carroll algebra has 6 generators: $\{L_0,L_1,L_{-1},M_0,M_1,M_{-1}$\}. It is expected that once we introduce a boundary, we obtain 3 independent non-zero symmetry generators. If we consider $\mathcal{O}$ and $\mathcal{P}$, this expectation is met and we indeed have 3 independent generators viz. $\{\mathcal{O}_1,\mathcal{P}_0,\mathcal{P}_1\}$. The algebra of these finite generators is given by 
\begin{align}
    [\mathcal{O}_1,\mathcal{P}_0]=2P_1,~~~[\mathcal{O}_1,\mathcal{P}_1]=2P_0,~~~[\mathcal{P}_0,\mathcal{P}_{1}]=0.  
\end{align}
This is actually the $iso(2)$ algebra (or $iso(1,1)$ since the algebra does not distinguish between spatial and temporal directions), as is to be expected. However, if we consider $\mathcal{O}$ and $\mathcal{R}$, $\mathcal{R}_0$ vanishes because antisymmetry, and we will only have $\{\mathcal{O}_1,\mathcal{R}_1\}$. These commute: 
\begin{align}
[\mathcal{O}_1,\mathcal{R}_1]~=~0.    
\end{align}
Given all of these reasons, going forward, we focus on $\{\mathcal{O}, \mathcal{P}\}$ as our bosonic BCCA generators.

\section{Carrollian Superconformal Field Theories}\label{CSCFT}
 Our first aim in this paper is to figure out how to supersymmetrize the above construction, i.e. we wish to build a supersymmetric boundary Carrollian CFT. As a prelude to this, we will begin this section by reviewing the supersymmetric extension of the 2d Carrollian CFTs. Just like the bosonic 2d CCFTs, these theories too, can be connected to appropriately defined Carrollian limits of the relativistic superconformal field theories (SCFTs). In earlier studies \cite{Bagchi:2016yyf,Bagchi:2017cte}, it was found that there are two possible ways to find supersymmetric extension of 2d CCA, namely Homogeneous and Inhomogeneous Superconformal Carroll Algebras (SCA). In this section, we review both of these algebras and also show how they come from different In\"on\"u-Wigner contractions of the Super-Virasoro algebra.

\medskip

We start from the supersymmetric extension of the Virasoro algebra. For SFT without boundaries we have symmetry generators $\{\mathcal{L}^+_{n},\mathcal{Q}_{r},\mathcal{L}^-_{n},\bar{\mathcal{Q}}_{r}\}$, which satisfy two copies of Super-Virasoro algebra given below
\begin{subequations}\label{sv2}
\begin{align}
    [\mathcal{L}^+_{n},\mathcal{L}^+_{m}]&=(n-m)\mathcal{L}^+_{n+m}+\frac{c}{12}(n^3-n)\delta_{n,-m}, \\
    [\mathcal{L}^+_{n},\mathcal{Q}_{r}]&=\Big(\frac{n}{2}-r\Big)\mathcal{Q}_{n+r},\\
    \{\mathcal{Q}_{r},\mathcal{Q}_{s}\}&=2\mathcal{L}^+_{r+s}+\frac{c}{3}\Big(r^2-\frac{1}{4}\Big)\delta_{r,-s}.
\end{align}
\end{subequations}
The similar algebra is satisfied by $(\mathcal{L}^-_n,\bar{\mathcal{Q}}_r)$ as well. There are two kinds of supersymmetric extensions of  Virasoro algebra taking the same form given in \eqref{sv2}. One is the Ramond algebra where $r\in\mathbb{Z}$ and the other is the Neveu-Schwarz algebra where  $r\in\mathbb{Z}+\frac{1}{2}$. 

\subsection{Homogeneous SSCA}
We will begin our Carrollian explorations with the so-called {\em Homogeneous} limit, where the supersymmetry generators would be scaled in the same way. We take
the following In\"on\"u-Wigner contraction on the symmetry generators 
\bea{homocon}
L_n=\mathcal{L}^+_n-\mathcal{L}_{-n}^-, \quad M_n = \eps (\mathcal{L}^+_n+\mathcal{L}^-_{-n}), \quad Q^+_r=\sqrt{\eps}\mathcal{Q}_r,\quad Q^-_r = \sqrt{\eps}\bar{\mathcal{Q}}_{-r}.
\eea
This is the homogeneous scaling since both $\mathcal{Q}^+_r$'s and $\mathcal{Q}^-_{r}$'s are being scaled by the same factor, $\sqrt{\epsilon}$. This scaling leads to the following algebra, which we call Homogeneous Superconformal Carrollian algebra (SCCA)
\bea{sgcah}
&& [L_n, L_m] = (n-m) L_{n+m} + \frac{c_L}{12} \, (n^3 -n) \delta_{n+m,0}, \nonumber\\
&& [L_n, M_m] = (n-m) M_{n+m} + \frac{c_M}{12} \, (n^3 -n) \delta_{n+m,0}, \\
&& [L_n, Q^\a_r] = \Big(\frac{n}{2} - r\Big) Q^\a_{n+r}, \quad \{Q^\a_r, Q^\b_s \} = \delta^{\a\b} \left[M_{r+s} + \frac{c_M}{6} \Big(r^2 - \frac{1}{4}\Big)  \delta_{r+s,0} \right], \nonumber
\eea
where $\alpha$ and $\beta$ stand for $\pm$. In the rest of the paper, the homogeneous algebra would be the main focus of our attention. So to make the notation less cumbersome, we will not put any $(H)$ labels on this. This above algebra SCCA turns out to be the asymptotic symmetry algebra of conventional $\mathcal{N}=2$ supergravity in 3d asymptotically flat spacetimes \cite{Barnich:2014cwa}.

\subsection{Inhomogeneous SCCA}
The contraction from the two copies of the Virasoro algebra to the conformal Carroll algebra involved a linear combination of the bosonic generators. It is thus natural to envision such linear combinations while scale the supersymmetric generators as well. Interestingly, this leads to a richer algebra which has more non-trivial commutation relations. To see this consider the following In\"on\"u-Wigner contraction of Super-Virasoro algebra \eqref{sv2}: 
\bea{ingenscal}
L_n&=&\mathcal{L}^+_n-\mathcal{L}^-_{-n}, \quad M_n = \eps (\mathcal{L}^+_n+\mathcal{L}^-_{-n}), \nonumber \\
G_r&=&\mathcal{Q}_r-i \Bar{\mathcal{Q}}_{-r}, \quad K_r = \eps (\mathcal{Q}_r+i\Bar{\mathcal{Q}}_{-r}).
\eea
This kind of scaling on the superconformal generators is known as inhomogeneous scaling since unlike the homogenous cases, $G$ and $K$ are contracted in two distinct ways. The algebra we get from this contraction is known as inhomogeneous super Carrollian conformal algebra (SCCA$_I$),
\bea{sgcai}
&& [L_n, L_m] = (n-m) L_{n+m} + \frac{c_L}{12} (n^3 -n) \delta_{n+m,0}, \nonumber\\
&& [L_n, M_m] = (n-m) M_{n+m} + \frac{c_M}{12} (n^3 -n) \delta_{n+m,0}, \\
&& [L_n, G_r] = \Big(\frac{n}{2} -r\Big) G_{n+r}, \ [L_n, K_r] = \Big(\frac{n}{2} -r\Big) K_{n+r}, \ [M_n, G_r] = \Big(\frac{n}{2} -r\Big) K_{n+r}, \nonumber\\
&& \{ G_r, G_s \} = 2 L_{r+s} + \frac{c_L}{3} \Big(r^2 - \frac{1}{4}\Big)   \delta_{r+s,0}, \ \{ G_r, K_s \} = 2 M_{r+s} + \frac{c_M}{3} \Big(r^2 - \frac{1}{4}\Big)   \delta_{r+s,0}.\nonumber
\eea
We see now that in contrast to the homogeneous version above, the supersymmetric generators ``square-up'' to both the bosonic generators, giving more non-trivial structures in the SUSY sector of the algebra. It is interesting to note that this algebra has been also realized as an asymptotic symmetry algebra of an exotic theory of supergravity in 3d asymptotically flat spacetimes \cite{Lodato:2016alv}. 

\medskip

We will be interested in applications to string theory, in particular to null superstrings. Both algebras have been realised in terms of worldsheet symmetries of closed null superstrings. The homogeneous one was first obtained many years ago in \cite{Bagchi:2016yyf} and then understood as a Carrollian algebra in \cite{Bagchi:2022owq}. The inhomogeneous one led to the discovery of a new null superstring in \cite{Bagchi:2017cte}. Its tensile origins were explored in \cite{Bagchi:2018wsn}. We will return to string theory at length in the second half of the paper.

\section{Putting boundaries in  Superconformal Carroll Field Theory}\label{BasantaBiswas}
As described earlier, we wish to study the effect of boundaries in Superconformal Carrollian field theories (SCCFT). We will adapt an algebraic approach here where we deal specifically with generators and not their vector field representations. When we will address string theory later in the paper, we will return to the explicit vector field representations in terms of superspace. 

\subsection{Homogeneous theory with boundaries}
We begin by exploring the Homogeneous SCCA in the presence of boundaries. In order to obtain the boundary version of the SCCA, we use the following basis of generators 
\begin{align}\label{combination1}
    \mathcal{O}_{n}&=L_n-L_{-n},~~~~\mathcal{P}_n=M_n+M_{-n},\quad 
    \mathcal{H}_{r}=Q^{+}_{r}+Q^{-}_{-r}, ~~~~
    \mathcal{I}_{r}=Q^{+}_{r}-Q^{-}_{-r}.
\end{align}
Additionally, we define
\begin{align}
 \mathcal{R}_{n}&=M_n-M_{-n},   
\end{align}
and note that this generator is {\em not} a part of the bosonic BCCA. The redefined generators \eqref{combination1} are very similar to what we did for the bosonic theory and one of the supersymmetry generators would be picked out by the algebra. In order to see this, let us work out the commutators for the above sets of generators. 
\begin{subequations}
    \begin{align}
        &[\mathcal{O}_n,\mathcal{O}_m]=(n-m)\mathcal{O}_{n+m}-(n+m)\mathcal{O}_{n-m}\\&
    [\mathcal{O}_n,\mathcal{P}_m]=(n-m)\mathcal{P}_{n+m}+(n+m)\mathcal{P}_{n-m}\\&
    [\mathcal{O}_r,\mathcal{H}_{s}]=\Big(\frac{r}{2}-s\Big)\mathcal{H}_{r+s}+\Big(\frac{r}{2}+s\Big)\mathcal{H}_{s-r}\\&
    [\mathcal{O}_r,\mathcal{I}_{s}]=\Big(\frac{r}{2}-s\Big)\mathcal{I}_{r+s}+\Big(\frac{r}{2}+s\Big)\mathcal{I}_{s-r}\\&
    \{\mathcal{H}_{r},\mathcal{H}_{s}\}=\mathcal{P}_{r+s},~~~~\{\mathcal{I}_{r},\mathcal{I}_{s}\}=\mathcal{P}_{r+s},\\&
    \textcolor{red}{\{\mathcal{H}_{r},\mathcal{I}_{s}\}=\mathcal{R}_{r+s}},~~~~[\mathcal{H}_{r},\mathcal{P}_s]=[\mathcal{P}_r,\mathcal{P}_s]=0.
    \end{align}
\end{subequations}
We draw the attention of the reader to the bracket between $\mathcal{H}_{r}$ and $\mathcal{I}_{s}$ (in red above). From this we see, when we choose $\mathcal{O}, \mathcal{P}$ to be our bosonic generators of the BCCA (as we have done above), we cannot choose both $\mathcal{H}_{r}$ and $\mathcal{I}_{s}$ as their supersymmetric completion. We would have to choose one, and interestingly these lead to isomorphic algebras. For our explicit computations in the later section, we will choose $\mathcal{H}_{r}$ and this is the one we display as the Boundary Superconformal Carroll Algebra (BSCCA), now displayed with central terms:   
\begin{equation}\label{tintin}
    \begin{split}
        &[\mathcal{O}_n,\mathcal{O}_m]=(n-m)\mathcal{O}_{n+m}-(n+m)\mathcal{O}_{n-m}\\
    &[\mathcal{O}_n,\mathcal{P}_m]=(n-m)\mathcal{P}_{n+m}+(n+m)\mathcal{P}_{n-m}+\frac{c_M}{12}(n^3-n)(\delta_{n+m}+\delta_{n-m})\\
    &[\mathcal{O}_r,\mathcal{H}_{s}]=\Big(\frac{r}{2}-s\Big)\mathcal{H}_{r+s}+\Big(\frac{r}{2}+s\Big)\mathcal{H}_{s-r}\\
    &\{\mathcal{H}_{r},\mathcal{H}_{s}\}=\mathcal{P}_{r+s}+\frac{c_M}{3}(r^2-\frac{1}{4})\delta_{r+s,0},~~~
    [\mathcal{H}_{r},\mathcal{P}_s]=[\mathcal{P}_r,\mathcal{P}_s]=0.\\
    \end{split}
\end{equation}
We will recover the same BSCCA as a residual gauge symmetry algebra of tensionless homogeneous open superstring worldsheet. In our superspace formulation, which we will introduce in the string theory section, once boundary is introduced in appropriate location of the superspace, between $\mathcal{H}$'s and $\mathcal{I}$'s, one set of the generators becomes incompatible with the boundary and hence need to be removed. For our choice of boundary in superspace, the generators $\mathcal{I}$'s are removed and we will be left with the algebra above. 

\subsection*{From relativistic to Carroll boundary superconformal symmetry}
We now show how the BSCCA emerges from the Carroll limit of a single copy of superconformal algebra of super Virasoro generators expressed in a different basis. Analogous to Virasoro generators (in different basis) we discussed in the previous section, here we define the super Virasoro generators as follows
\begin{subequations}\label{Arjunda}
    \begin{align}
         &\mathbb{L}_{n}= \mathcal{L}^+_{n}+\mathcal{L}^-_{n}, \quad \widetilde{\mathbb{L}}_{n}= \mathcal{L}^+_{n}-\mathcal{L}^-_{n}, \quad \mathbb{Q}_{r}=\mathcal{Q}_{r}+\bar{\mathcal{Q}}_{r}, \quad \widetilde{\mathbb{Q}}_{r}=\mathcal{Q}_{r}-\bar{\mathcal{Q}}_{r}.
    \end{align}
\end{subequations}
The super-Virasoro algebra in this basis becomes following
\begin{subequations}\label{svbold}
    \begin{align}
        [\mathbb{L}_{n},\mathbb{L}_{m}]&=(n-m)\mathbb{L}_{(m+n)}~~~[\mathbb{L}_{n},\mathbb{Q}_{r}]=\Big(\frac{n}{2}-r\Big)\mathbb{Q}_{n+r}~~~\{\mathbb{Q}_{r},\mathbb{Q}_{s}\}=2\mathbb{L}_{r+s}\\
        [\widetilde{\mathbb{L}}_{n},\widetilde{\mathbb{L}}_{m}]&=(n-m)\mathbb{L}_{(m+n)}~~~[\widetilde{\mathbb{L}}_{n},\widetilde{\mathbb{Q}}_{r}]=\Big(\frac{n}{2}-r\Big)\mathbb{Q}_{n+r}~~~\{\widetilde{\mathbb{Q}}_{r},\widetilde{\mathbb{Q}}_{s}\}=2\mathbb{L}_{r+s}\\
        [\mathbb{L}_{n},\widetilde{\mathbb{L}}_{m}]&=(n-m)\widetilde{\mathbb{L}}_{n+m}~~~~~[\mathbb{L}_{n},\widetilde{\mathbb{Q}}_{r}]=\Big(\frac{n}{2}-r\Big)\widetilde{\mathbb{Q}}_{n+r}~~~\{\mathbb{Q}_{r},\widetilde{\mathbb{Q}}_{s}\}=2\widetilde{\mathbb{L}}_{r+s}.
    \end{align}
\end{subequations}
We have argued in our bosonic review that for boundaries placed at 
$(\sigma=0,\pi)$, $\widetilde{\mathbb{L}}_n$'s are not compatible generators as they contain $\partial_\sigma$ terms. The boundary preserving bosonic algebra thus just contains only $\mathbb{L}$. It is clear from the cross-commutator between $\mathbb{Q}$'s and $\widetilde{\mathbb{Q}}$'s, one set of the generators becomes incompatible with the boundary and hence needs to be removed.
At the level of the algebra, this choice is arbitrary and results in identical boundary algebras like in the case discussed above. We discard generators $\widetilde{\mathbb{Q}}$'s, and obtain an algebra containing ($\mathbb{L},\mathbb{Q}$). 
\begin{align}\label{bsv}
    [\mathbb{L}_{n},\mathbb{L}_{m}]&=(n-m)\mathbb{L}_{(m+n)}~~~[\mathbb{L}_{n},\mathbb{Q}_{r}]=\Big(\frac{n}{2}-r\Big)\mathbb{Q}_{n+r}~~~\{\mathbb{Q}_{r},\mathbb{Q}_{s}\}=2\mathbb{L}_{r+s}.
\end{align}
This choice will later be justified by our superspace analysis.  From the above algebra, one can obtain SCCA \eqref{tintin} through the following In\"on\"u-Wigner contraction
\begin{align}\label{HIWC}
    \mathcal{O}_n=\mathbb{L}_n-\mathbb{L}_{-n},~~~P_n=\epsilon~(\mathbb{L}_n+\mathbb{L}_{-n}),~~~\mathcal{H}_
    r=\sqrt{\epsilon}\mathbb{Q}_{r},~~~\e\to 0.
\end{align}
It is interesting to ask whether the above is the only BSCCA that one could obtain from the Homogeneous SCA. From our algebraic analysis above, it is clear that when the bosonic sub-algebra is fixed to be the BCCA \eqref{bdyCarr}, and when we consider the Homogeneous supersymmetric extension, \eqref{tintin} is the unique algebra that one can obtain. The choice between $\mathcal{H}$'s and $\mathcal{I}$'s does not change the algebra. 

\subsection{Inhomogeneous SCCA with boundaries}
We now turn our attention to the Inhomogeneous algebra \eqref{sgcai}. We will play the same game as in the Homogeneous case. To begin with we choose the following combinations of generators: 
\begin{subequations}\label{inhom-comb}
\begin{align}
\mathcal{O}_n&=L_n-L_{-n},~~~\mathcal{P}_n=M_n+M_{-n}\\
\mathcal{J}_r&=G_r-G_{-r},\quad \mathcal{K}_r=K_r+K_{-r} \quad \mathcal{G}_r=G_r+G_{-r},~~~\mathcal{Y}_r=K_r-K_{-r}
\end{align} 
\end{subequations}
We also take into account other generators:
\begin{align}
   \mathcal{X}_n&=L_n+L_{-n},~~~\mathcal{R}_n=M_n-M_{-n} 
\end{align}
We note that these $(\mathcal{X}, \mathcal{R})$ are {\em not} generators of the bosonic BCCA \eqref{bdyCarr}. We now write down the algebra of the generators \eqref{inhom-comb}, omitting the commutators of bosonic generators given in \eqref{bdyCarr} and zero brackets: 
\begin{align}
    &\left[\mathcal{O}_n, Z_r\right]=\left(\frac{n}{2}-r\right)Z_{n+r}-\left(\frac{n}{2}+r\right)Z_{n-r}, \quad \text{where} \quad Z \equiv \mathcal{J}, \mathcal{G}, \mathcal{K}, \mathcal{Y}.\nonumber\\
    &\left[\mathcal{P}_n,\mathcal{J}_r\right]=\left(\frac{n}{2}-r\right)\mathcal{K}_{n+r}-\left(\frac{n}{2}+r\right)\mathcal{K}_{n-r}, \nonumber\\ &\left[\mathcal{P}_n,\mathcal{G}_r\right]=\left(\frac{n}{2}-r\right)\mathcal{Y}_{n+r}+\left(\frac{n}{2}+r\right)\mathcal{Y}_{n-r} \\
    &\textcolor{red}{\left\{\mathcal{J}_r,\mathcal{J}_s\right\}=2\mathcal{X}_{n+r}-2\mathcal{X}_{n-r},}\quad \textcolor{red}{\left\{\mathcal{G}_r,\mathcal{G}_s\right\}=2\mathcal{X}_{n+r}+2\mathcal{X}_{n-r},} \nonumber \\&\left\{\mathcal{J}_r,\mathcal{G}_s\right\}=2\mathcal{O}_{n+r}-2\mathcal{O}_{n-r},\quad \textcolor{red}{\left\{\mathcal{K}_r,\mathcal{J}_s\right\}=2\mathcal{R}_{n+r}-2\mathcal{R}_{n-r},} \nonumber \\&\left\{\mathcal{K}_r,\mathcal{G}_s\right\}=2\mathcal{P}_{n+r}+2\mathcal{P}_{n-r}\quad \left\{\mathcal{Y}_r,\mathcal{J}_s\right\}=2\mathcal{P}_{n+r}-2\mathcal{P}_{n-r}\quad \textcolor{red}{\left\{\mathcal{Y}_r,\mathcal{G}_s\right\}=2\mathcal{R}_{n+r}+2\mathcal{R}_{n-r}} \nonumber.
\end{align}
These generators $(\mathcal{O}, \mathcal{P}, \mathcal{J}, \mathcal{K}, \mathcal{G}, \mathcal{Y})$ {\em do not} form a closed algebra. The problematic brackets are pointed out in red. It is clear that we would not be able to include $\mathcal{J}_r$ and $\mathcal{G}_r$ in a set of generators that contain generators of BCCA \eqref{bdyCarr} as the only bosonic generators. So only the permissible generators will be $\mathcal{O}_n$, $\mathcal{P}_n$, $\mathcal{K}_r$, and $\mathcal{Y}_r$, which now close to form an inhomogeneous version of BSCCA that we regard as BSCCA$_I$,
\begin{subequations}\label{proredin}
\begin{align}
    &\left[\mathcal{O}_n,\mathcal{O}_m\right]=(n-m)\mathcal{O}_{n+m}-(n+m)\mathcal{O}_{n-m}\\
    &\left[\mathcal{O}_n,\mathcal{P}_m\right]=(n-m)\mathcal{P}_{n+m}+(n+m)\mathcal{P}_{n-m}+\frac{c_M}{12}(n^3-n)(\delta_{n+m}+\delta_{n-m})\\
    &\left[\mathcal{O}_n,\mathcal{K}_r\right]=\left(\frac{n}{2}-r\right)\mathcal{K}_{n+r}+\left(\frac{n}{2}+r\right)\mathcal{K}_{n-r}\\&\left[\mathcal{O}_n,\mathcal{Y}_r\right]=\left(\frac{n}{2}-r\right)\mathcal{Y}_{n+r}-\left(\frac{n}{2}+r\right)\mathcal{Y}_{n-r}\\&
\{\mathcal{K}_r,\mathcal{K}_s\} = \{\mathcal{Y}_r,\mathcal{Y}_s\}
\;= \{\mathcal{Y}_r,\mathcal{K}_s\}
\;=\;
[\mathcal{P}_r,\mathcal{K}_s] = [\mathcal{P}_r,\mathcal{Y}_s]
\;=\;
[\mathcal{P}_n,\mathcal{P}_m] = 0,
\end{align}
\end{subequations}
where we have now also included central terms. The only non-zero central term is in the bosonic sector. The above algebra is somewhat unimpressive in the sense that the fermionic generators don't do much. They don't even ``square'' to the bosonic generators, -- they just anticommute.  
From a particular In\"on\"u-Wigner contraction of \eqref{bsv}, one can obtain the algebra given in \eqref{proredin}. The contraction is as follows.
\begin{align}\label{Prolimin}
    &\mathcal{O}_n=\mathbb{L}_n-\mathbb{L}_{-n},~~~P_n=\epsilon~(\mathbb{L}_n+\mathbb{L}_{-n}),\\&\mathcal{K}_
    r=\epsilon(\mathbb{Q}_{r}+\mathbb{Q}_{-r}),~~~~\mathcal{Y}_
    r=\epsilon(\mathbb{Q}_{r}-\mathbb{Q}_{-r}),~~~\e\to 0.
\end{align}
The algebra \eqref{proredin}, if restricted to one of the two generators $\mathcal{K}_r$ or $\mathcal{Y}_r$ is also obtainable as a further contraction of the Homogeneous algebra \eqref{tintin}. So the boundary version of the Inhomogeneous SCCA is actually less rich than the Homogeneous version, in direct contrast to the original algebras without the introduction of boundaries. 

\medskip

In what follows, we will be interested in the Homogeneous SCA and will ignore the inhomogeneous version. It is possible that when we look at a different type of boundary in the original 2d bosonic CCFT where instead of $\mathcal{O}$, we use $\mathcal{X}$ e.g. for a boundary in $\tau$ instead of $\sigma$, the inhomogeneous version of the supersymmetric theory would become more prominent. We leave this for future explorations. 

\medskip

This concludes our generic exploration of Supersymmetric BCCFTs in this paper. We will now focus on a particular application of our general results in this part: the appearance of the Homogeneous version of the BSCCA as the residual symmetries on the worldsheet of the null open string. 

\newpage

\section{Bosonic Null strings}\label{bosonic null strings}

As stressed in the introduction, null tensionless strings are to string theory what massless point particles are to usual physics based on point particles. Just as massless particles travel along null geodesics in spacetime, null strings sweep out 2d null worldsheets in the target space. 

\medskip

The recent revival of null strings have been focused on the understanding of the ILST action \cite{Isberg:1993av} and the emergent 2d Carrollian conformal structures that appear on the null worldsheet \cite{Bagchi:2013bga, Bagchi:2015nca, Bagchi:2020fpr, Bagchi:2019cay, Bagchi:2020ats}. The null superstrings in the same way has intimate connections with the 2d Super Carroll CFTs, - both the homogeneous \cite{Bagchi:2016yyf} and inhomogeneous versions \cite{Bagchi:2017cte} we encountered in the previous sections. 

\medskip

The recent analyses of null strings have been focused on the closed string sector and it was an open and challenging question what open strings meant in this context \footnote{See previous attempts in \cite{Sagnotti:2011jdy, Sagnotti:2003qa,Bonelli:2003kh, Lindstrom:2003mg,Bakas:2004jq, Francia:2007qt}.}. This question was answered in \cite{Bagchi:2024qsb} and it was found that the Boundary CCFT emerged as the worldsheet residual symmetry on the bosonic open null worldsheet. 

\medskip

We now ask how to formulate open superstrings in the null or tensionless limit. This will be the focus of this part of the paper. For the sake of completeness and for better readability, we will walk the reader through the basic construction of open bosonic null strings. We then delve into the details of the open null superstring and show how the BSCCA emerges on the supersymmetric null open worldsheet. In Appendix \ref{tensile superstring}, we provide a concise description of open tensile superstrings with a particular focus on the superspace formulation, which will be useful for our constructions of the open null superstring.

\subsection{Bosonic null strings: Closed}

The point-particle limit of string theory is the limit where the length of the fundamental string $\ell_s=2\pi \alpha'=1/T$ goes to zero and this reduces string theory to Einstein gravity. Here $T$ is the tension of the fundamental string. The other diametrically opposite limit is where the tension of the string goes to zero $T\to0$ making the fundamental string long, floppy and tensionless. It is this rather strange limit we would be interested in. As mentioned earlier, this is analogous to the massless limit of the point particle and the worldsheet becomes null.  

\medskip

The worldsheet formulation of null strings was first developed in \cite{Lindstrom:1990qb} and later rigorously studied in \cite{Isberg:1993av}. The worldsheet dynamics of null strings are governed by the ILST action, which takes the form on a 2d manifold $\mathcal{M} (\tau, \sigma)$,
\begin{equation}
    S_{\text{ILST}} = \frac{1}{2 \pi c'}\int_{\mathcal{M}} d\tau~d\sigma\,~ V^\alpha V^\beta(\partial_\alpha X^\mu)(\partial_\beta X^\nu) \eta_{\mu\nu},
    \label{eq:ILST}
\end{equation}
 where $c'$ is a dimensionful normalization constant, and it plays the role analogous to $\alpha'$ in the standard tensile theory. Once we vary the ILST action with respect to $X^\mu$ and $V^\alpha$, we obtain the following equations of motion and the constraint equations,
\begin{eqnarray}
\partial_\alpha(V^\alpha V^\beta\partial_\beta X^\nu)\eta_{\mu\nu}=0, ~~
V^\beta\eta_{\mu\nu}(\partial_\alpha X^\mu)(\partial_\beta X^\nu) = 0,
\label{eq:eom}
\end{eqnarray}
together with boundary terms. For the moment, we focus on closed null strings with periodic boundary conditions. We will come back to the boundary terms in detail when we discuss open null strings.  

\medskip
 
As in the tensile case for the Polyakov action, the ILST action has Poincaré invariance in target spacetime as well as diffeomorphism invariance ($\xi^\alpha\to\xi^{\alpha}+\epsilon^{\alpha}$) on the worldsheet. Here, under diffeomorphism, $X^{\mu}$ and $V^{\alpha}$ transform in the following way
\begin{align}\label{B3}
   \delta V^\alpha=-&V^\beta\partial_\beta\epsilon^\alpha+\epsilon^\beta\partial_\beta V^\alpha+\frac{1}{2}(\partial_\beta\epsilon^\beta)V^\alpha,\quad 
\delta X^{\mu}(\tau,\sigma)=\epsilon^{\alpha}\partial_{\alpha}X^{\mu}(\tau,\sigma). 
\end{align}
As the worldsheet diffeomorphism is a gauge symmetry, the null string action (\ref {eq:ILST}) requires gauge fixing. A convenient choice of gauge is the following
\begin{align}\label{Review1}
    V^{\alpha}=(v,0), \hspace{5mm}v=\text{constant}.
\end{align}
It can be shown that in this gauge, there is a residual symmetry left over, again very much like the tensile string, but the usual two copies of the Virasoro algebra that form the worldsheet symmetry for the usual string theory now on the null worldsheet is replaced by the 2d Conformal Carroll algebra \eqref{bms3}. The methods of 2d CCFTs have been used to understand null bosonic closed strings \cite{Bagchi:2013bga, Bagchi:2015nca, Bagchi:2020fpr, Bagchi:2019cay, Bagchi:2020ats}. For details of these constructions, the reader is referred to \cite{Bagchi:2015nca}. 

\subsection{Bosonic null strings: Open}
We now return to the boundary term which we neglected in the variation of the ILST action. This is given by 
\begin{equation}
\int_{\partial\cal M}\extd\tau\,n_\alpha V^\alpha V^\beta(\partial_\beta X^\nu)\eta_{\mu\nu}\,\delta X^\mu = 0.
\label{eq:beom}
\end{equation}
Here $\partial \mathcal{M}$ is the spatial boundary located at $\sigma = 0, \pi$ (as shown in the figure \ref{fig2}) and $n_\alpha$ is normal to the boundary. In order to make the boundary term vanish, we have the following options on boundary conditions at $\sigma=0,\pi$
\begin{subequations}
\label{unusual}
\begin{align}
    \delta X^\mu \big|_{\sigma=0,\pi}&=0 \qquad \text{Dirichlet,}\label{eq:Dirichlet}\\ 
    V^\beta\partial_\beta X^\nu \big|_{\sigma=0,\pi}&=0 \qquad \text{Neumann,}\\
    n_\alpha V^\alpha\big|_{\sigma=0,\pi} &=0 \qquad \text{Null.} \label{eq:DerDritteMann}
    \end{align}
\end{subequations}
The first and second conditions, respectively, correspond to Dirichlet and Neumann boundary conditions analogous to the tensile open string. The third condition \eqref{eq:DerDritteMann}, however, particularly appears for the null strings only. It is particularly intriguing that in the conformal gauge we are interested in \eqref{Review1}, this condition is automatically satisfied. This hints to a formal understanding of the folklore about open and closed strings becoming identical in the tensionless limit. 

\medskip

For what follows, here and in the supersymmetric case later, we will focus on the Dirichlet boundary condition. The other two, especially the null boundary condition as we described above, are very interesting and we hope to return to them in the near future.  

\medskip

We will be using the gauge \eqref{Review1} even here. To understand the symmetries on the worldsheet, now we adopt a slightly different route. The equations of motion and the constraint equations for the gauge fixed  action are
\begin{subequations}\label{tsc3}
\begin{align}
    \ddot {X}^\mu=0,\\
\label{constraintsboson}
   \dot{X}^2=0,\hspace{5mm}\dot{X}\cdot X'=0.
   \end{align}
\end{subequations}
The most general solution to the equation of motion $\ddot {X}^\mu=0$ becomes
\begin{align}
\label{genmode1}
    X^{\mu}(\tau,\sigma)=x^{\mu}+\sqrt{\tfrac{c'}{2}}\,(C^{\mu}_{0}&-\Tilde{C}^{\mu}_{0})\,\sigma+\sqrt{\tfrac{c'}{2}}\,(C^{\mu}_{0}+\Tilde{C}^{\mu}_{0})\,\tau \nonumber\\&+i\sqrt{\tfrac{c'}{2}}\sum_{n\neq0}\frac{e^{-in\sigma}}{n}\left[(C^{\mu}_{n}-\Tilde{C}^{\mu}_{-n})-in\tau(C^{\mu}_{n}+\Tilde{C}^{\mu}_{-n})\right],
\end{align}
where $C^{\mu}_n$ and $\tilde{C}^{\mu}_n$ are oscillator modes (i.e. they obey usual simple harmonic oscillator commutators, more below). Imposing the Dirichlet boundary condition on (\ref{genmode1}) at $\sigma = 0, \pi$ yields $ C^\mu_n = - \tilde{C}^\mu_n$ and thus we obtain the mode expansion for the null open string \footnote{Once we impose the periodic boundary condition $X^{\mu}(\tau, \sigma+2 \pi) = X^{\mu}(\tau, \sigma)$ on (\ref{genmode1}) we obtain the mode expansion of null closed strings \cite{Bagchi:2015nca}.}, 
\begin{align}
    X^{\mu}(\tau,\sigma)=x_0^{\mu}+\sqrt{2c'}&C^{\mu}_{0}\sigma +i\sqrt{\frac{c'}{2}}\sum_{n\neq0}\frac{1}{n}\big[(C^{\mu}_{n}+C^{\mu}_{-n}) -in\tau(C^{\mu}_{n}-C^{\mu}_{-n})\big]e^{-in\sigma},
    \label{modeexpansion1}
\end{align}
where the fact that $(C_{n}^\mu)^{\dagger}=C_{-n}^\mu$ ensures that the modes are real. Here we introduce the zero mode as
\begin{equation}
C_0^\mu=(x_1^\mu-x_0^\mu)/(\sqrt{2c^\prime}\pi),
\end{equation}
where $X^{\mu} (0)= x_{0}^{\mu}$, $X^{\mu} (\pi)=x_{1}^{\mu}$.
Once we insert the mode expansion of the null open strings (\ref{modeexpansion1}) to compute $\dot X^\mu, X'^\mu$ and use the result in (\ref{constraintsboson}), the constraint equations translate into quadratic relations in terms of those corresponding null open string oscillator modes.  

\begin{subequations}\label{eq:con}
\begin{align}
&\dot{X}^2=\sum_{m} \mathcal{P}_m e^{-im\sigma} = 0, \\
& \dot{X}\cdot{X}^\prime = \sum_{m}\big(\mathcal{O}_m-im\tau \mathcal{P}_m\big)e^{-im\sigma}=0\,, 
\end{align}
\end{subequations}
where we define
\begin{subequations}
    \label{eq:bi}
\begin{align}
 \mathcal{P}_n&=\sum_{m}\frac{1}{4}(C_m^\mu-C_{-m}^\mu)(C_{n-m}^\nu-C_{m-n}^\nu)\eta_{\mu\nu},\\
\mathcal{O}_{m}&=\sum_{n}\frac{1}{2}(C_n^\mu-C_{-n}^\mu)(C_{m-n}^\nu+C_{n-m}^\nu)\eta_{\mu\nu}.
\end{align}
\end{subequations}
In order to quantize the null open string, we need to promote the Poisson bracket structure between the canonically conjugate variables $X^\mu$ and $ P^{\mu} = \frac{1}{2 \pi c'}\dot{X}^\nu$
into the commutator bracket of the respective operators and 
the classical constraints~\eqref{eq:con} 
are elevated to operator conditions that act on physical states and are required to vanish. The commutator bracket between $X^\mu$ and $P^{\mu}$ leads to,
\begin{equation}
[C_n^\nu,\,C_m^\mu] = n\,\delta_{n,-m}\,\eta^{\mu\nu}\,.
\label{eq:osc}
\end{equation}
Using the above commutation relations \eqref{eq:osc}, it is straightforward to show that the constraint generators $\mathcal{O}_m$ and $\mathcal{P}_m$ satisfy the Boundary Carrollian Conformal Algebra \eqref{bdyCarr}  shown earlier from the field theoretic arguments. It is important to note that the same BCCA algebra can be independently recovered from the residual gauge symmetry analysis after imposing the gauge (\ref{Review1})  in the presence of the Dirichlet boundary condition imposed at $\sigma = 0, \pi$ \cite{Bagchi:2024qsb}.

 \section{Null superstrings}\label{null superstrings}
In this section, we construct a theory of open null superstrings based on the Dirichlet conditions advertised above. It has already been shown in the tensionless version of closed string theory that there exists two independent versions of the tensionless sector of the closed superstring, e.g., the homogeneous \cite{Bagchi:2016yyf} and the inhomogeneous \cite{Bagchi:2017cte} tensionless superstrings. Similarly, we naturally expect these two independent versions of tensionless superstring to still persist in the presence of a boundary. For our present purpose, we focus on the homogeneous version of the tensionless open superstring satisfying the Dirichlet boundary condition. In our previous analysis of the symmetry algebra, we saw that the homogeneous version led to a more interesting boundary algebra and we will realise this algebra in the context of string theory in the following. 

\subsection{Closed null superstrings}

The tensionless superstring action is given by \cite{Lindstrom:1990ar}
\begin{equation}\label{hcfer1}
S=\int d^{2} \xi ~~ \left[\left(V^{\a}\partial_{\a} X^{\mu}+i \chi \psi^{\mu}\right) \cdot\left(V^{\b} \partial_{\b} X^{\mu}+i \chi \bar{\psi}^{\mu}\right)+i\bar\psi^{\mu}\rho^{\a}  \partial_{\a}\psi_{\mu}\right].
\end{equation}
In this action $\psi^{\mu}$ is the fermionic density carrying a weight $-\frac{1}{4}$, and pairs with the bosonic partner $X^{\mu}$. The fermionic partner for vector density $V^{\a}$ is $\chi$, where the latter is related to the ordinary gravitino $\chi_{\a}$ by the relation $\chi=V^\a\chi_{\a}$. Once we vary the above action for supersymmetric null strings with respect to $X^\mu$, $\psi_\mu$, and $\chi_\alpha$ we obtain the respective equations of motion
\begin{subequations}
\begin{align}\label{hcfer2}
\partial_{\a}\left(V^{\a} V^{\b} \partial_{\b} X_{\mu}\right)+i \partial_{\a}\left(V^{\a} V^{\b} \chi_{\b} \psi_{\mu}\right)- i \partial_{\a}\left(\rho^{\a} \psi_{\mu}\right)=0, \\
2 V^{\a} V^{\b}\left(i \chi_{\b} \partial_{\a} X_{\mu}+\chi_{\a} \chi_{\b} \psi_{\mu}\right)=0,\\
 V^{\b}\left(\partial_{\a} X^{\mu} \partial_{\b} X_{\mu}+2 i \chi_{\b} \psi^{\mu} \partial_{\a} X_{\mu}-\chi_{\a} \psi^{\mu} \chi_{\b} \psi_{\mu}\right)+\frac{i}{2} \psi^{\mu} \partial_{\a} \psi_{\mu}=0,  \\
 V^{\a} V^{\b} \psi^{\mu}\left(i \partial_{\b} X_{\mu}-\chi_{\b} \psi_{\mu}\right)=0,
\end{align}
\end{subequations}
and a boundary term that we shall revisit shortly when we introduce null supersymmetric open strings.
The action \eqref{hcfer1} is invariant under worldsheet diffeomorphisms and supersymmetry transformations. Under the  worldsheet diffeomorphism parametrized by $\xi^{\a}$ the fields vary as 
\begin{subequations}
\begin{align}\label{chhagol}
\delta_{\xi} X^{\mu} & =-\xi^{\b} \partial_{\b} X^{\mu},~~~~
\delta_{\xi} V^{\a} =-V^{\b} \partial_{\b} \xi^{\a}+\xi^{\b} \partial_{\b} V^{\a}+\frac{1}{2}\left(\partial_{\b} \xi^{\b}\right) V^{\a},\\
\delta_{\xi}\psi^{\mu} & =-\xi^{\b} \partial_{\b} \psi^{\mu},~~~~
\delta_{\xi} \chi  =-\xi^{\a} \partial_{\a} \chi+\frac{1}{4}\left(\partial_{\a} \xi^{\a}\right) \chi.
\end{align}
\end{subequations}
Supersymmetry transformation among the fields with parameter $\epsilon$ produces
\begin{subequations}\label{hcfer4}
\begin{align}
\delta_{\epsilon} X^{\mu} & =i \epsilon \psi^{\mu},~~~~\delta_{\epsilon} \psi^{\mu}  =-\epsilon V^{\a} \partial_{\a} X^{\mu}-\frac{1}{2} i \epsilon\left(\psi^{\mu} \chi\right) , \\ \label{chhagol1}
\delta_{\epsilon} V^{\a}& =i V^{\a}\left(\epsilon \chi\right),~~~~
\delta_{\epsilon} \chi =V^{\a} \partial_{\a} \epsilon .
\end{align}
\end{subequations}
The presence of the diffeomorphism on the null superstring worldsheet, like in the bosonic case, leads to a gauge fixing condition. This will be the tensionless equivalent of the conformal gauge for the usual superstring: 
\begin{equation}
\label{gauge}
V^{\a}=(1,0),~~~~~~~ \chi=0. 
\end{equation}
Note that the gamma matrices in the tensile theory satisfy the usual form of Clifford algebra \eqref{Cliff} where we use $\eta^{ab}$ as the gauge fixed tensile worldsheet metric. However, here, due to the presence of degenerate metric structure on the null worldsheet, we replace $\eta^{\a\b}$ by an appropriate combination of the gauged fixed vector densities.  With this modification, the Clifford algebra of the gamma matrices in the tensionless superstring theory reads 
\begin{eqnarray}
\{\rho^{\a}, \rho^\b\} = 2 V^\a V^\b .
\label{Clifftless}
\end{eqnarray}
A convenient representation of this modified Clifford algebra \eqref{Clifftless} is obtained by simply identifying the gamma matrices  $\rho^\a$  with the vector densities $V^\a$, i.e., $\rho^\a = V^\a \mathbb{1}$ where the explicit form of the $\rho^\a$ matrices for homogeneous null closed superstrings is given as 
\begin{eqnarray}
\label{rho}
\rho^0 = \mathbb{1}, ~~\rho^1 = \mathbb{O}.
\end{eqnarray}
In this gauge, the action becomes:
\begin{equation}\label{hcfer5}
S=\int d^{2} \xi\left[\dot{X}^{2}+i \psi \cdot \dot{\psi}\right].
\end{equation}
The equations of motion and the constraints take the following form 
\begin{subequations} \label{hcfer7}
\begin{align}\label{hcfer71}
&\ddot{X}^{\mu}=0, \quad \dot{\psi}^{\mu}=0,\\ \label{hcfer72}
&\dot{X}^{2}=0, \quad \dot{X} \cdot X^{\prime}+\frac{i}{2} \bar{\psi} \cdot \psi^{\prime}=0, \quad \psi \cdot \dot{X}=0, 
\end{align}
\end{subequations}
and the gauge fixed forms of the restricted diffeomorphism and the supersymmetry
under which the action \eqref{hcfer5} remains invariant are given as 
\begin{subequations}\label{homosusy}
\begin{align}
\delta_\xi X&=\xi^\a\p_\a X, \quad \delta_\xi \psi=\xi^\a\p_\a \psi +\frac{1}{4}(\p_\a\xi^\a)\psi, \\
\delta_\e X&=\bar{\e}\psi, \quad \delta_\e \psi^\mu=-i\rho^\a\p_\a X \e.
\end{align}
\end{subequations}
Using the gauge in the second equations of \eqref{chhagol} and \eqref{chhagol1}, we obtain the  following differential equations for $\xi$ and $\epsilon$ :
\begin{equation}\label{hcfer9}
\dot{\xi^{0}}=\left(\xi^{1}\right)^{\prime}, \quad \dot{\xi^{1}}=0, \quad \dot{\epsilon^{\alpha}}=0.
\end{equation}
Solving the above equation gives us the form for the restricted diffeomorphism and supersymmetry transformations:
\begin{equation}\label{hcfer10}
\xi^{0}=f^{\prime}(\sigma) \tau+g(\sigma), \quad \xi^{1}=f(\sigma), \quad \epsilon^{ \pm}=\epsilon^{ \pm}(\sigma).
\end{equation}
To clearly illustrate the form of the generators for  \eqref{hcfer10}, we need to extend those transformations of $\xi$ and $\epsilon$ to $\mathcal{N}=(1,1)$ superspace by adjoining Grassmann coordinates $\theta$ and $\bar{\theta}$ to the two‑dimensional world‑sheet coordinates $\sigma^{\a}$. This uplift yields the following transformations on the homogeneous Carrollian superspace \cite{Bagchi:2016yyf},
\begin{subequations}\label{carrollsupertrans}
  \begin{align}
\sigma^{\a} & \rightarrow \sigma^{\prime a}=\sigma^{\a}+\xi^{\a}+\frac{i}{2} \epsilon^{+} V^{\a} \theta+\frac{i}{2} \epsilon^{-} V^{\a} \bar{\theta}  \\
\theta & \rightarrow \theta^{\prime}=\theta+\epsilon^{+}+\frac{1}{4} \theta \partial_{\a} \xi^{\a}  \\
\bar{\theta} & \rightarrow \bar{\theta}^{\prime}=\bar{\theta}+\epsilon^{-}+\frac{1}{4} \bar{\theta} \partial_{\a} \xi^{\a} 
\end{align}  
\end{subequations}
Consequently, \eqref{carrollsupertrans} induce the following transformation of a superfield $\Phi\left(\sigma^{\a}, \theta, \bar{\theta}\right)$, 
\begin{equation}\label{kumropotash}
\delta \Phi=\left(\delta \tau \partial_{\tau}+\delta \sigma \partial_{\sigma}+\delta \theta \partial_{\theta}+\delta \bar{\theta} \partial_{\bar{\theta}}\right) \Phi 
\end{equation}
By making use of  \eqref {gauge}, \eqref{rho}, \eqref{hcfer10} and \eqref{carrollsupertrans} in \eqref{kumropotash} we obtain an explicit form of the superfield transformation, 
\begin{align}
\delta \Phi= & {\left[\left(f^{\prime} \tau+g+\frac{i}{2} \epsilon^{+} \theta+\frac{i}{2} \epsilon^{-} \bar{\theta}\right) \partial_{\tau}+f \partial_{\sigma}\right.} \nonumber\\
&~~~~~~~~~~~~~~~~~~ \left.+\left(\epsilon^{+}+\frac{1}{2} f^{\prime} \theta\right) \partial_{\theta}+\left(\epsilon^{-}+\frac{1}{2} f^{\prime} \bar{\theta}\right) \partial_{\bar{\theta}}\right] \Phi  \nonumber\\
= & {\left[L(f)+M(g)+Q^{+}\left(\epsilon^{+}\right)+Q^{-}\left(\epsilon^{-}\right)\right] \Phi } .
\end{align}
Once we have this explicit form, it becomes straightforward to identify the generators of superfield transformation,
\begin{subequations}
\label{cgen}
\begin{align}
&L(f)  =f \partial_{\sigma}+f^{\prime}\left[\tau \partial_{\tau}+\frac{1}{2}\left(\theta \partial_{\theta}+\bar{\theta} \partial_{\bar{\theta}}\right)\right] , \quad M(g)=g \partial_{\tau},  \\
Q^{+}&\left(\epsilon^{+}\right) =\epsilon^{+}\left(\partial_{\theta}+\frac{i}{2} \theta \partial_{\tau}\right), \quad Q^{-}\left(\epsilon^{-}\right)=\epsilon^{-}\left(\partial_{\bar{\theta}}+\frac{i}{2} \bar{\theta} \partial_{\tau}\right), 
\end{align}
\end{subequations}
where $L$ and $M$ are the superrotation and supertranslation generators in the bosonic sector, while $Q^{ \pm}$ are fermionic generators. The parameters $f, g$ and $\epsilon^{ \pm}$ are functions of $\sigma$. Since they are periodic functions of sigma (due to the periodicity of the closed string worldsheet), they can be Fourier expanded as
\begin{equation}
\label{genmode2}
f=\sum_{n} a_{n} e^{i n \sigma}, \quad g=\sum_{n} b_{n} e^{i n \sigma}, \quad \epsilon^{ \pm}=\sum_{n} \zeta_{n}^{ \pm} e^{i n \sigma}.
\end{equation}
Once we substitute \eqref{genmode2} back into \eqref{cgen}, we re-express the bosonic and fermionic generators in the corresponding modes, 
\begin{equation}
L(f)=-i \sum_{n} a_{n} L_{n}, \quad M(g)=-i \sum_{n} b_{n} M_{n}, \quad Q^{ \pm}\left(\epsilon^{ \pm}\right)=\sum_{n} \zeta_{n}^{ \pm} Q_{n}^{ \pm},
\end{equation}
where the mode operator $L_n$ and $M_n$ are given as , 
\begin{subequations}\label{HSBMS}
    \begin{align}
&L_n=ie^{in\sigma}\Big[\partial_\sigma+in\tau\partial_\tau+\frac{in}{2}\big(\theta\partial_\theta+\bar{\theta}\partial_{\bar{\theta}}\big)\Big],~~~M_n=ie^{in\sigma}\partial_\tau\\
    &Q^{+}_{r}=e^{ir\sigma}\Big(\partial_{\theta}+\frac{i}{2}\theta\partial_{\tau}\Big),~~~~Q^{-}_{r}=e^{ir\sigma}\Big(\partial_{\bar{\theta}}+\frac{i}{2}\bar{\theta}\partial_{\tau}\Big).
    \end{align}
\end{subequations}
A direct computation of the commutators and anticommutators among $L_n$, $M_n$ and $Q^{\pm}$ by using their operator forms confirms that they close into the homogeneous super‑Carrollian conformal algebra (SCCA), excluding the central term.
\begin{equation}\label{HSBMSAlgebra}
    \begin{split}
       &[L_n,L_m]=(n-m)L_{n+m}~~~~[L_n,M_m]=(n-m)M_{n+m}\\
        &[L_n,Q^{\alpha}_{r}]=\Big(\frac{n}{2}-r\Big)Q^\alpha_{n+r}~~~\{Q^{\alpha}_{r},Q^{\alpha'}_{s}\}=\delta^{\alpha\alpha'}M_{r+s}\\
       &[M_m,M_n]=[M_m,Q^{\alpha}_{r}]=0.
   \end{split}
\end{equation}
 In the quantum theory of the null closed strings, we can recover the homogeneous SCCA algebra, including the central terms derived earlier in \eqref{sgcah} from field theoretic arguments. 

\medskip

\subsection{Open null superstrings: Boundary conditions}
In this section, we present for the first time, a construction of an open null superstring satisfying Dirichlet boundary conditions and demonstrate how the null open worldsheet intrinsically realizes boundary homogeneous superconformal  Carrollian algebra (BSCCA).
Here, we primarily focus on the boundary term, that we previously did not examine when we discussed the variation of the action \eqref{hcfer1}. 
\begin{equation}
\label{varaction}
\begin{split}
\delta S_{bdy} = -\frac{T}{2}\int d^2\sigma\, \Big[\,\partial_{\a}\,& \left(V^\a\, V^\b\,\partial_{\b} X_{\mu}\delta X^{\mu}+iV^\a V^\b\chi_{\b}\psi_\mu\delta X^{\mu}\right)  -i\partial_\a\left(\bar{\psi}^\mu \rho^\a\delta \psi_\mu\right)\Big].
\end{split}
\end{equation}

The purely fermionic part of the boundary term given by the last term in \eqref{varaction} can be recast into \begin{eqnarray}
\label{fbc}
\int d^2\sigma \,\partial_\a\left(\bar{\psi}^\mu \rho^\a\,\delta \psi_\mu\right) = \oint_c dl ~ (\,\bar{\psi}^\mu\, n_\a \rho^\a\,\delta\psi_\mu\,).
\end{eqnarray}
One way to make this term vanish is the fermionic analogue of the null boundary conditions \eqref{eq:DerDritteMann} we found in the bosonic case:
\begin{eqnarray}
n_\a \rho^\a = 0.
\end{eqnarray}
Although we will not venture in this direction in the current work, this is clearly an interesting avenue to pursue. We plan to return to this in future work. 

For the present purpose, our interest lies only in the Dirichlet boundary conditions, and we will work with the same gauge as mentioned in \eqref{gauge}. The representation of the gamma matrices stays the same as \eqref{rho}. Additionally,  for this particular choice of gauge, the action, equations of motion, constraint equations, restricted diffeomorphism and the supersymmetry transformation for the homogeneous null open superstrings remain the same as \eqref{hcfer5}, \eqref{hcfer7}, and \eqref{homosusy} respectively. Interestingly, choice of the gauge $V^a=(1,0)$ or \{$\rho^{0} = \mathbb{1}, \rho^1 = \mathbb{O}$\}  enforces the  fermionic boundary term given in \eqref{fbc} to vanish identically.
\begin{align}
 \oint_c dl ~ n_a (\,\bar{\psi}\,\rho^a\,\delta\psi\,) = \int d\sigma\,( \psi_+\,\delta\psi_++\,\psi_-\,\delta\psi_-)\Big|_{\tau=-\infty}^{\tau=\infty}=0,
 \end{align}

Unlike the tensile superstrings, the way the fermionic boundary term becomes zero does not lead to any particular boundary condition. Nonetheless, we still have a choice to introduce the Dirichlet boundary conditions in the fermionic sector by combining the bosonic Dirichlet condition with the gauge‑fixed supersymmetry transformations \eqref{homosusy}.

\begin{equation}\label{jalebi}
\delta_\e X\Big|_{\s=0,\pi}=0, \qquad \implies \epsilon^+\,\psi_+=-\,\epsilon^-\,\psi_-.
\end{equation}
Note that the coefficients of $\psi_+$ and $\psi_-$ given above differ from those in the tensile case \eqref{rasagolla} due to the distinct representation of gamma matrices. Similar to the theory of tensile superstrings, here also we exploit the fact that the presence of the boundary breaks half of the supersymmetry and thus following the similar steps given in \eqref{sec2.1}, we obtain both  Neveu-Schwarz as well as Ramond sector of the homogeneous null open superstrings satisfying the the Dirichlet boundary condition. The Neveu-Schwarz sector of the fermionic field becomes
\begin{subequations} \label{tensionlessns}
\begin{align}
    & \epsilon^+(0)\,= \epsilon^-(0)\, \implies \psi_+(0)=-\,\psi_-(0),\\
    & \epsilon^+(\pi)= -\epsilon^-(\pi)\, \implies \psi_+(\pi)=\psi_-(\pi).
    \end{align}
\end{subequations}
Likewise, the Ramond sector takes the following form
\begin{subequations}\label{tensionlessramond}
\begin{align}
    \epsilon^+(0)\,= \epsilon^-(0)\, &\implies \psi_+(0)=-\psi_-(0),\\
    \epsilon^+(\pi)= \epsilon^-(\pi)\, &\implies \psi_+(\pi)=-\psi_-(\pi).
    \end{align}
\end{subequations}
The above two equations \eqref{tensionlessns} and \eqref{tensionlessramond} give us the necessary constraints on the fermionic fields which we use to study their mode expansions etc. For the rest of our analysis we will be focussing on the Neveu-Schwarz sector.

\subsection{Symmetry generators with superspace boundary}\label{hijibijbij2}
It is well known that in tensile RNS superstring theory, the gauge fixed action can also be rewritten 
on a superspace where there are worldsheet coordinates \{$\t,\s$\} and Grassman coordinates \{$\theta,\bar{\theta}$\}. Hence, for a tensile open superstring, the boundaries introduced on the worldsheet at $\s=0,\pi$ should also have a superspace extension.
The superspace boundary locations for both Neveu Schwarz and Ramond sectors for tensile RNS superstrings have been worked out for Dirichlet boundary conditions in Appendix \ref{dinosaur}. For Dirichlet Neveu Schwarz case, the superspace boundary was found to be located at  
\begin{align}\label{susybdy}
\{\sigma = 0, \theta_2 = 0\} \quad \text{and} \quad \{\sigma = \pi, \theta_1=0\}, \quad \text{where} \quad \theta_{1,2} = \frac{1}{2}(\theta \pm \bar{\theta}). 
\end{align}
In this section, we discuss the fate of the homogeneous super CCA$_2$ generators in the presence of the same superspace boundaries \eqref{susybdy}. 
Here we have assumed that the boundary locations in the superspace are not affected by the homogeneous Carrollian limit on the superspace. The assumptions will be justified once we look into the limiting study of this theory. Note that our choice of boundary is restricted to the Dirichlet Neveu Schwarz sector of the open null homogenous superstring.
We know from \cite{Bagchi:2024qsb}, the  CCA$_2$ symmetry generators in the presence of boundary ($\sigma = 0$ and $\sigma = \pi$) reduces to the following boundary compatible generators
\begin{align}\label{combination}
    \mathcal{O}_{n}=L_n-L_{-n},~~~~\mathcal{P}_n=M_n+M_{-n}.
\end{align}
Hence, we need to ensure whether this combination of generators also preserves the aforementioned superspace boundary. The generators $L_n$s and $M_n$s as given in \eqref{HSBMS} will lead us to the vector field expressions:  
    \begin{align}\label{pteranodon}
    \mathcal{O}_{n}=-2\sin{n\sigma}\partial_{\sigma}-2n\tau\cos{n\sigma}\partial_{\tau}-n\cos{n\sigma}(\theta_1\partial_{\theta_1}+\theta_2\partial_{\theta_2}),\quad 
    \mathcal{P}_{n}=2i\cos{n\sigma}\partial_{\tau}.
    \end{align}
As one can see, none of the generators has any $\partial_\sigma$ or $\partial_{\theta_2}$ term at $\{\sigma=0, \theta_2=0\}$. Similarly, at the other boundary at $\{\sigma=\pi, \theta_1=0\}$ all the terms containing $\partial_\sigma$ or $\partial_{\theta_1}$ disappear. 

So none of these generators distort the boundary \eqref{susybdy} and hence for SCCA too, \eqref{combination} is the correct way to find the BSCCA generators. Now, let us look into the generators $Q_{r}^{\pm}$. These generators can be rewritten in a new basis
\begin{subequations}\label{T-Rex}
\begin{align}
    \mathcal{H}_{r}&=Q^{+}_{r}+Q^{-}_{-r}=\cos{r\sigma}(\partial_{\theta_1}+i\theta_{1}\partial_\tau)+i\sin{r\sigma}(\partial_{\theta_2}+i\theta_2\partial_\tau),\\
    \mathcal{I}_{r}&=Q^{+}_{r}-Q^{-}_{-r}=i\sin{r\sigma}(\partial_{\theta_1}+i\theta_{1}\partial_\tau)+\cos{r\sigma}(\partial_{\theta_2}+i\theta_2\partial_\tau).
\end{align}
\end{subequations}
Now, inspecting $\mathcal{H}_{r}$'s and $\mathcal{I}_{r}$'s one can immediately see that no $\partial_{\sigma}$ term is appearing in the generators $\mathcal{H}_{r}$'s and any term containing $\partial_{\theta_2}$  vanishes at the boundary \{$\sigma = 0, \theta_2 = 0$\}. However, a term with $\partial_{\theta_2}$ still remains in the generators $\mathcal{I}_{r}$'s. Moreover, since  $n\in\mathbb{Z}$ but $r\in\mathbb{Z}+\frac{1}{2}$ in the NS sector of null homogeneous superstring, no term associated with $\partial_{\theta_1}$ survives in $\mathcal{H}_{r}$'s at the other boundary located at $\sigma=\pi$, $\theta_1=0$.
This confirms that $\mathcal{H}_{r}$'s are legitimate symmetries for  superspace boundary defined at \{$\sigma = 0, \theta_2 = 0$\} and \{$\sigma = \pi, \theta_1=0$\}.
Hence, we are left with only the generators $\mathcal{O}_n, \mathcal{P}_n$ and $\mathcal{H}_{r}$, and the complete symmetry we have is given below
\begin{subequations}\label{BSBMS}
    \begin{align}
        \mathcal{O}_{n}&=-2\sin{n\sigma}\partial_{\sigma}-2n\tau\cos{n\sigma}\partial_{\tau}-n\cos{n\sigma}(\theta_1\partial_{\theta_1}+\theta_2\partial_{\theta_2})\\
        \mathcal{P}_{n}&=2i\cos{n\sigma}\partial_{\tau}\\
         \mathcal{H}_{r}&=\cos{r\sigma}(\partial_{\theta_1}+i\theta_{1}\partial_\tau)+i\sin{r\sigma}(\partial_{\theta_2}+i\theta_2\partial_\tau).
    \end{align}
\end{subequations}
These generators $\mathcal{O}_n, \mathcal{P}_n$ and $\mathcal{H}_{r}$ close under the following algebra
\begin{equation}\label{tintin1}
    \begin{split}
        &[\mathcal{O}_n,\mathcal{O}_m]=(n-m)\mathcal{O}_{n+m}-(n+m)\mathcal{O}_{n-m}\\
    &[\mathcal{O}_n,P_m]=(n-m)P_{n+m}+(n+m)P_{n-m}\\
    &[\mathcal{O}_r,\mathcal{H}_{s}]=\Big(\frac{r}{2}-s\Big)\mathcal{H}_{r+s}+\Big(\frac{r}{2}+s\Big)\mathcal{H}_{s-r}\\
    &\{\mathcal{H}_{r},\mathcal{H}_{s}\}=\mathcal{P}_{r+s}~~~~[\mathcal{H}_{r},\mathcal{P}_s]=[\mathcal{P}_r,\mathcal{P}_s]=0.
    \end{split}
\end{equation}
We regard this algebra as the boundary version of the homogeneous super CCA$_2$ algebra. The same algebra, along with the corresponding central terms, was derived earlier from a field-theoretic analysis in equation \eqref{tintin}.

\subsection{Open null superstrings: Mode expansion and constraint algebra}

Now we come up with the mode expansion of the Dirichlet Neveu Schwarz sector of the null open homogeneous superstring and derive the boundary version of the homogeneous SCCA algebra from the constraint analysis.

We begin with the gauge-fixed equations of motion \eqref{hcfer71} where the bosonic field admits the same mode expansion as given in \eqref{genmode1}, while the fermionic fields follow the expansion given below
\begin{align}
\label{tlessmode}
     \psi^\mu_+(\tau,\sigma)=\sqrt{c'}\sum_{r\in\mathbb{Z}+\frac{1}{2}}\beta^{~\mu+}_{r}e^{-ir\sigma},\quad     \psi^\mu_-(\tau,\sigma)=\sqrt{c'}\sum_{r\in\mathbb{Z}+\frac{1}{2}}\beta^{~\mu-}_{r}e^{-ir\sigma}.
\end{align}
With the help of the bosonic and the fermionic Dirichlet boundary conditions as previously described in \eqref{eq:Dirichlet} and in \eqref{tensionlessns} respectively, we identify the modes in the bosonic sector as well as in the Neveu Schwarz fermionic sectors as, 
\begin{align}\label{strongconstraint}
    C_n^\mu&=-\tilde{C}_n^\mu ~~\forall n, \quad
     \beta^{~\mu+}_{r}=-\beta^{~\mu-}_{-r} ~~\forall r.
    \end{align}
With these above identification, the bosonic sector reduces to the expansion given in \eqref{modeexpansion1}, while the fermionic sector takes the following form:
 \begin{align} \label{modeexpansion}
    \psi^\mu_+(\tau,\sigma)=&\sqrt{c'}\sum_{r\in\mathbb{Z}+\frac{1}{2}}\beta^{~\mu+}_{r}e^{-ir\sigma},\quad \psi^\mu_-(\tau,\sigma)=-\sqrt{c'}\sum_{r\in\mathbb{Z}+\frac{1}{2}}\beta^{~\mu+}_{-r}e^{-ir\sigma}.
\end{align}
The commutation relations follow by considering
\be{modecom}
[X^\mu(\sigma),\dot{X}^\nu(\sigma')]=\eta^{\mu\nu}\delta(\sigma-\sigma'), \quad \{ \psi^\mu_{+}(\sigma),\psi^\nu_{+}(\sigma') \}=\eta^{\mu\nu}\delta(\sigma-\sigma') ,
\ee
and that of the modes,
\be{abb}
[C_m^\mu,C_n^\nu] = 2m\delta_{m+n}\eta^{\mu\nu}, \quad \{\beta_{r}^{\mu+},\beta_{s}^{\nu+} \} = \delta_{r+s}\eta^{\mu\nu}.
\ee

We now use the mode expansions of the null open superstring, with the bosonic modes given in equation \eqref{genmode1} and the fermionic modes in equation \eqref{tlessmode}, and substitute them into the supersymmetric constraint equations given in \eqref{hcfer72} to obtain
\begin{subequations}
\label{m}
\begin{align}
\frac{1}{2}\,\dot X^2& =  \frac{c'}{4}\sum_{n,p}\left(C_n^\mu-C_{-n}^\mu\right)\left(C_{p-n}^\nu-C_{n-p}^\nu\right)\eta_{\mu\nu}e^{-ip\sigma}=c'\sum_{p} \mathcal{P}_pe^{-ip\sigma}\\
\dot X \cdot X'+\frac{i}{2}\,\bar{\psi}\cdot\psi' &= \frac{c'}{2}\sum_{p}\bigg[\sum_{n}\big(C_n^\mu-C_{-n}^\mu\big)\big(C_{p-n}^\nu+C_{n-p}^\nu\big)\nonumber\\&~~~~-i\,p\,\tau\,\frac{1}{2}\sum_{n}\big(C_n^\mu-C_{-n}^\mu)(C_{p-n}^\nu-C_{n-p}^\nu\big)\\&~~~~+\frac{1}{2}\sum_{r\in\mathbb{Z}+\frac{1}{2}}\,(2r+p)\,(\beta^{\mu+}_{-r}\beta^{\nu+}_{r+n})\,-(2r-p)\,(\beta^{\mu+}_{-r}\beta^{\nu+}_{r-n})\,\bigg]\eta_{\mu\nu}e^{-ip\sigma},\nonumber\\&=c'\sum_{p}\left(\mathcal{O}_p-i\,p\,\tau \mathcal{P}_p\right)e^{-ip\sigma}\nonumber \\
       \psi_+\cdot\dot{X} =\psi_-\cdot\dot{X}&=\frac{c'}{\sqrt{2}}\sum_{n,r}\big(C_n^\mu-C_{-n}^\mu\big)\,\beta^{\nu+}_{r+n}\,\eta_{\mu\nu}e^{-ir\sigma}     =c'\sum_{r\in\mathbb{Z}+\frac{1}{2}}\mathcal{H}_{r}e^{-ir\sigma}.
    \end{align}
\end{subequations}

Generators in terms of modes are given as,
\begin{subequations}\label{genmode}
\begin{align}
\mathcal{O}_{p} =~&  \frac{1}{2}\sum_{n}(C_n^\mu - C_{-n}^\mu)(C_{p-n}^\nu + C_{n-p}^\nu) \notag \\
&~~~~~~ + \frac{1}{4} \sum_{r\in\mathbb{Z}+\frac{1}{2}} \left[ (2r + p)\,(\beta^{\mu+}_{-r} \beta^{\nu+}_{r+n}) - (2r - p)\,(\beta^{\mu+}_{-r} \beta^{\nu+}_{r-n}) \right] \eta_{\mu\nu} \\
\mathcal{P}_p =~&  \frac{1}{4}\sum_{n}(C_n^\mu - C_{-n}^\mu)(C_{p-n}^\nu - C_{n-p}^\nu)\, \eta_{\mu\nu} \\
\mathcal{H}_{r} =~& \frac{1}{\sqrt{2}} \sum_{n} (C_n^\mu - C_{-n}^\mu)\, \beta^{\nu+}_{r+n} \eta_{\mu\nu}
\end{align}
\end{subequations}
In \eqref{genmode} we expressed  the generators $\mathcal{O}_n, \mathcal{P}_n$ and $\mathcal{H}_r$ in terms of the oscillator modes of the bosonic and fermionic sectors of the null open superstring.  Using the commutation and anticommutation relations satisfied by these bosonic and fermionic modes as mentioned in \eqref{abb}, we recover the homogeneous BSCCA \eqref{tintin1}.

\subsection{Null open superstring from limit}
Up to this point, our analysis has treated the tensionless superstring as an intrinsic object, allowing us to derive its key properties without involving any contraction of the world‑sheet. We now turn to the limiting approach, wherein we uncover the relation between the tensile open superstring and the homogeneous version of the tensionless open superstring by performing a Carroll contraction of the tensile string worldsheet. In order to do so, we begin with the tensile open superstring satisfying the Dirichlet boundary conditions. The appropriate mode expansions for bosonic field $X^{\mu}(\tau,\sigma)$ and the Neveu Schwarz sector of the fermionic field $\psi^{\mu}(\tau,\sigma)$ are given as,
\begin{subequations}
\label{tensi0}
\begin{align}
     X^{\mu}(\tau,\sigma)=x^{\mu}+\sqrt{2\alpha'}\alpha_0^\mu\s+i\sqrt{\frac{\alpha'}{2}}\sum_{n\neq0}\frac{1}{n}\left[\alpha^{\mu}_{n} e^{-in(\tau+\sigma)} + \alpha^{\mu}_{-n}e^{in(\tau-\sigma)}\right],\\ \label{tensi1}\psi^{\mu}_{+}(\tau,\sigma)=\sqrt{\alpha'}\sum_{r\in\mathbb{Z}+\frac{1}{2}}b^\mu_re^{-ir(\tau+\sigma)},\quad \psi^\mu_{-}(\tau,\sigma)=-\sqrt{\alpha'}\sum_{r\in\mathbb{Z}+\frac{1}{2}}b^\mu_re^{-ir(\tau-\sigma)},
     \end{align}
\end{subequations}
where $\alpha^{\mu}_n$ and $b^{\mu}_r$ satisfy the following brackets.
\begin{align}\label{mode}
[\a_m^\mu,\a_n^\nu]=m\delta_{m+n}\eta^{\mu\nu},\qquad
\{b_{r}^{\mu},b_{s}^{\nu} \}=\delta^{\a\a'}\delta_{r+s}\eta^{\mu\nu}.
\end{align} 
From the constraint equations, we obtain the components of the energy-momentum tensor and the supercurrents for the tensile supersymmetric string, as given below
\begin{subequations}\label{constraints}
\begin{align}
    T_{++}&=(\partial_{+}X)^2+\frac{i}{2}\psi_{+}\cdot\partial_{+}\psi_{+}=0,\qquad T_{--}=(\partial_{-}X)^2+\frac{i}{2}\psi_{-}\cdot\partial_{-}\psi_{-}=0\\ &   ~~~~~~~J_{+}=\psi_{+}\cdot\partial_{+}X=0,\qquad J_{-}=\psi_{-}\cdot\partial_{-}X=0.
\end{align}
\end{subequations}
The super-Virasoro generators for the open strings (at $\tau=0$) are as follows
\begin{subequations}
    \begin{align}
        \mathbb{L}_n&=\frac{1}{2\pi \a'}\int_0^\pi d\s \left(e^{in\s}\, T_{++} + e^{-in\s}\, T_{--}\right),\\
        \mathbb{Q}_r&=\frac{1}{2\pi\a'}\int_0^{\pi}d\s\, (e^{in\s}  J_+ +e^{-in\s} J_{-}).
    \end{align}
\end{subequations}
In terms of oscillator modes, these super-Virasoro generators take the form
\begin{subequations}\label{supvir}
\begin{align}
    &\mathbb{L}_n=\frac{1}{2}\sum_m \alpha_{-m}\cdot \alpha_{m+n}+\frac{1}{4}\sum_{r\in\mathbb{Z}+\frac{1}{2}} (2r+n)b_{-r}\cdot b_{r+n},\\
    &\mathbb{Q}_{r}=\frac{1}{\sqrt{2}}\sum_{m}\alpha_{-m}\cdot b_{r+m}.
\end{align}
\end{subequations}
These generators satisfy the super-Virasoro algebra, as given in \eqref{bsv}, which was derived from field-theoretic considerations.

\medskip

We now proceed to obtain the homogeneous version of the tensionless open superstring from the tensile open superstring discussed above, by performing a limiting analysis. The tensionless limit on the worldsheet is realized through an ultra-relativistic or Carroll scaling of the worldsheet coordinates: 
\begin{align}
    \tau\to\epsilon\tau,~~\alpha'\to\frac{c'}{\epsilon}~~\sigma\to\sigma,~~\epsilon\to0.
\end{align}
In addition, it is essential to specify the scaling behaviour of the fermionic fields in the tensionless superstring limit. Given that the underlying symmetry is governed by the homogeneous BSCCA, the fermions scale homogeneously as 
\begin{align}
    \psi_{\text{tensionless}} = \sqrt{\epsilon} \, \psi_{\text{tensile}}.
\end{align}
Applying this scaling on \eqref{tensi0} we obtain,
\begin{subequations}
\begin{align}
\label{tensi0.1}
     X^{\mu}(\tau,\sigma)&=x^{\mu}+\sqrt{\frac{2c'}{\epsilon}}\alpha_0^\mu\sigma+i\sqrt{\frac{c'}{2\epsilon}}\sum_{n\neq0}\frac{1}{n}\Big[\alpha^{\mu}_{n} (1-i\epsilon n\tau) + \alpha^{\mu}_{-n}(1+i\epsilon n\tau)\Big]e^{-in\sigma},\\\label{tensi2}
     \psi^\mu_{+}(\tau,\sigma)&=\sqrt{\e\alpha'}\sum_{r\in\mathbb{Z}+\frac{1}{2}}b^\mu_re^{-ir\sigma}(1-i\epsilon r\tau)\approx\sqrt{c'}\sum_{r\in\mathbb{Z}+\frac{1}{2}}b^\mu_re^{-ir\sigma}+\mathcal{O}(\epsilon)\\ \label{tensi21}
    \psi^\mu_{-}(\tau,\sigma)&=-\sqrt{\e\alpha'}\sum_{r\in\mathbb{Z}+\frac{1}{2}}b^\mu_re^{ir\sigma}(1-i\epsilon r\tau)\approx-\sqrt{c'}\sum_{r\in\mathbb{Z}+\frac{1}{2}}b^\mu_{-r}e^{-ir\sigma}+\mathcal{O}(\epsilon)
\end{align}
\end{subequations}
Now, as we  compare the scaled modes \eqref{tensi0} with the intrinsic modes in the bosonic sector \eqref{modeexpansion1} and also in the fermionic sector \eqref{modeexpansion}, we achieve the following relations,  
 \begin{subequations}    \label{openbv}
\begin{align}
C^{\mu}_{n}&=\frac{1}{2}\Big(\sqrt{\epsilon}+\frac{1}{\sqrt{\epsilon}}\Big)\alpha^{\mu}_{n}-\frac{1}{2}\Big(\sqrt{\epsilon}-\frac{1}{\sqrt{\epsilon}}\Big){\alpha}^{\mu}_{-n},\\ \label{beta}
    \beta^{\mu+}_{r}&=b^\mu_r.
\end{align}
\end{subequations}
Plugging \eqref{openbv} into \eqref{genmode}, we obtain the following expressions for the homogeneous tensionless superstring generators in terms of tensile modes
\begin{subequations}\label{tensilegen}
\begin{align}
\mathcal{O}_{n} =~&  \frac{1}{2}\sum_{p}(\alpha_p\cdot\alpha_{n-p}-\alpha_{-p}\cdot\alpha_{p-n}) \notag \\
&~~~~~~ + \frac{1}{4} \sum_{r\in\mathbb{Z}+\frac{1}{2}} \big[ (2r + p)\,(b^{\mu}_{-r}\, b^{\nu}_{r+n}) - (2r - p)\,(b^{\mu}_{-r}\, b^{\nu}_{r-n}) \big] \eta_{\mu\nu}, \\
\mathcal{P}_n =~&  \frac{\epsilon}{4}\sum_{p} \left(\alpha_p\cdot\alpha_{n-p}+\alpha_{-p}\cdot\alpha_{p-n}\right), \\
\mathcal{H}_{r} =~& \sqrt{\frac{\e}{2}} \sum_{n} (\a_n^\mu - \a_{-n}^\mu)\, b^{\nu}_{r+n} \eta_{\mu\nu}.
\end{align}
\end{subequations}
Using the oscillator algebra of the open string modes $\alpha^{\mu}_n$ and $b^{\mu}_r$ given in \eqref{mode}, it can be readily shown that the homogeneous tensionless superstring generators $\mathcal{O}_n$, $\mathcal{P}_n$ and $\mathcal{H}_r$ expressed in the form of bilinear operators in \eqref{tensilegen}
indeed satisfy the BSCCA without central terms as previously mentioned in \eqref{tintin1}. Also for consistency, expressing \eqref{tensilegen} in terms of the super-Virasoro generators \eqref{supvir}, we recover the relations between the generators of the tensile and tensionless open superstrings that were obtained from field-theoretic considerations in \eqref{HIWC}. Hence, this limiting analysis explicitly establishes the emergence of the boundary realization of the homogeneous super Carrollian conformal algebra, presented in \eqref{tintin1}.

\subsection*{Limit From Boundary super Virasoro algebra}\label{Hijibijbij}
As a final sanity check, we will perform a limit from the vector field representations of the Super Virasoro in superspace to the BSCCA.  
The vector field form of the super Virasoro generators $\mathbb{L}_{n}$ and $\mathbb{Q}_r$ which are derived in appendix \ref{tensile superstring} (see \eqref{different1} and \eqref{different3})
compatible with the superspace boundary condition \eqref{susybdy}.  
\begin{equation}\label{bscft}
    \begin{split}
        \mathbb{L}_{n}&=ie^{in\sigma^+}\Big(\partial_++\frac{in}{2}\theta\partial_\theta\Big)+ie^{in\sigma^-}\Big(\partial_-+\frac{in}{2}\Bar{\theta}\partial_{\Bar{\theta}}\Big)\\
         \mathbb{Q}_{r}&=e^{ir\sigma^+}(\partial_{\theta}+i\theta\partial_+)+
e^{ir\sigma^-}(\partial_{\Bar{\theta}}+i\Bar{\theta}\partial_-).
    \end{split}
\end{equation}
Homogeneous scaling of the super Virasoro generators is introduced by the following scaling of the superspace coordinates, 
\begin{align}
    \tau\to \epsilon\tau,
~~~~\sigma\to\sigma,~~~~
\theta_1\to\sqrt{\epsilon}\theta_1,~~~~\theta_2\to\sqrt{\epsilon}\theta_2, 
\end{align}
where $\theta = \theta_1+\theta_2,~\bar{\theta} = \theta_1-\theta_2$ \footnote{From the limiting perspective, we can justify that, the location of the superspace boundaries in tensile open superstring theory is still valid in the homogeneous null open superstring theory. The homogeneous scaling on the Grassmannian coordinates $\theta_1$ and $\theta_2$ in the tensile theory does not affect our analysis, since we only zoom around $\theta_1 \,= \,\theta_2 = 0$.}.
Applying this scaling on the superspace coordinates \eqref{bscft}, we can perform the In\"onu-Wigner contraction on the Boundary Super-Virasoro generators given in \eqref{HIWC}. The coordinate representations of $\mathcal{O}_n$'s, $\mathcal{P}_{n}$'s and $\mathcal{H}_{r}$'s obtained in this manner turn out to be exactly the same as those obtained in an intrinsic manner in \eqref{BSBMS}. This confirms our earlier intrinsic analysis.

\section{Conclusion and future directions}\label{conclusions}

\subsection*{Summary}
The current paper could be broadly divided in two parts. In the first part of this paper, constituting of Sec.~2 to 4, we formulated how to put boundaries on Supersymmetric Carrollian CFTs in $d=2$. Taking cues out of the bosonic version of the construction in \cite{Bagchi:2024qsb}, which we recapitulated in Sec.~2, we explored how one can restrict the boundary algebra in the supersymmetric case. The SUSY extension already had two versions of Carroll conformal algebras, viz. the homogeneous and inhomogeneous (which we revisited in Sec.~3), and we found boundary versions in both. We showed how to obtain these algebras from a single copy of the Super Virasoro algebra. There were some surprises along the way. The inhomogeneous algebra, which is richer in terms of the number of non-zero commutators between generators, turned out to be less interesting when boundaries were inserted than the homogeneous version. We explained this phenomenon algebraically.  

\medskip

The second part of the paper dealt with string theory and in particular the first construction of a open null superstring. We again gave the reader a quick tour of the basics of closed null strings and their SUSY extension, as well as the construction of the bosonic open null string. The null open superstring was constructed by combining the constructions of the null bosonic open string and the null closed superstring. We concentrated on the homogeneous version and found that the same algebra as discovered in the first half of the paper was reproduced in the context of null open superstings thus providing a very satisfying first application of the new algebra. 

\subsection*{Looking ahead}
There are several pressing questions which arise from our analysis and ones we hope to solve in the near future. Below we list a few of them.
\begin{itemize}
 \item {\em Inhomogeneous open superstrings:} We have seen that there are two boundary superconformal Carroll algebras arising from the homogeneous and inhomogeneous algebras. We dealt with the homogeneous algebra when we constructed the null superstring. This is the open string equivalent of the construction in \cite{Bagchi:2016yyf}. There exists an inhomogeneous construction for the null superstring \cite{Bagchi:2017cte}. It seems that the open sector of this superstring is somewhat less interesting and more restrictive from the point of view of the algebra which is like a contracted version of the homogeneous boundary algebra. It would be very interesting to figure out why there is such a difference in the homogeneous and inhomogeneous sectors of the open null superstring. 
    
\item {\em Null boundary conditions:} In our analysis of boundary terms for the open null strings, we came across a rather curious condition which we called the null boundary condition \eqref{eq:DerDritteMann}. In the conformal gauge we were interested in for both the bosonic and the supersymmetric case, we saw that this condition was automatically satisfied hence there was no need to impose any further boundary condition from the point of view of the intrinsic null string analysis. Our construction in this paper (as also in the original \cite{Bagchi:2024qsb}) focused however on the Dirichlet condition. We did this in order to connect with the boundary Carroll CFT constructed in the first half of the paper and also to tie up closely with the open tensile string. The theory constructed for the null open (super)string was a direct descendant of the tensile parent open (super)string. The null boundary condition, as mentioned earlier, is an avenue for further exploration into the belief that open and closed strings lose their distinction in the tensionless limit. This gives further credence to the novel closed to open transition encountered in \cite{Bagchi:2019cay}.

\item {\em D-branes:} We have looked at a theory of open strings and the natural next question is what of D-branes. Should all D-branes constructed out to tensionless theories be necessarily null branes or is there some wiggle-room for non-null branes? We hope to be able to use techniques in relativistic boundary CFTs like the construction of boundary states to understand D-branes from the point of view of worldsheet symmetries. 

\item {\em Quantization:} The quantization of the closed bosonic null string led to three different quantum theories built on three different vacua, which were the induced, flipped and oscillator vacua \cite{Bagchi:2020fpr}. It would be interesting to understand whether all of these vacua support open strings and if all types of boundary conditions are compatible with these different vacua and hence different theories. The equivalent superstring analysis has not been performed and is a very important step to take. 

\item {\em Open-closed dualities:} The re-examination of open-closed dualities for the null string also remains an intriguing question. This is especially so as there is also the transition from closed to open strings as tension goes to zero \cite{Bagchi:2019cay}. 
\end{itemize}

We hope to address the above questions as well as several others regarding the field theory aspects of Boundary CCFTs, with and without supersymmetry, in the near future.

\bigskip

\subsection*{Acknowledgements}
We thank Daniel Grumiller and Stefan Fredenhagen for discussions about boundary CCFTs and for collaboration on the bosonic ancestor of this paper \cite{Bagchi:2024qsb}. 

\medskip
AB is partially supported by a Swarnajayanti Fellowship from the Science and Engineering Research Board (SERB) under grant SB/SJF/2019-20/08. AB, SC and RC acknowledge support from ANRF grant CRG/2022/006165. PP acknowledges support from the Infosys Endowment for the Study of the Quantum Structure of Spacetime, and earlier from  IIT Kanpur Institute Assistantship for Postdoctoral Fellow. Most of this work was carried out at IIT Kanpur, whose hospitality is gratefully acknowledged, along with that of the University of Edinburgh, University of Southampton, TIFR, and ICTS-TIFR. PP thanks the organisers of the “DAE-BRNS Symposium” at IIT BHU and “Non-Lorentzian Geometries and Their Applications” at the Hamilton Mathematics Institute, Trinity College Dublin, for their hospitality. RC and PP thank the organisers and participants of “Strings 2025” at New York University Abu Dhabi, for useful discussions and warm hospitality. Finally, PP thank the organisers and participants of the “National Strings Meet (NSM)” at IIT Ropar for their hospitality and stimulating discussions.

\newpage

\appendix 

\section*{APPENDICES}

\section{Review of Basics: Tensile open superstring}\label{tensile superstring}
In this Appendix we review the tensile open superstring with an emphasis on its superspace formalism.  
We begin with the Ramond-Neveu-Schwarz (RNS) action in superconformal gauge for a supersymmetric tensile string 
\be{taction}
S=-\frac{T}{2}\int d^2\sigma \Big[\p_aX^\mu\p^aX_\mu-i\bar{\psi}^\mu\rho^a\p_a\psi_\mu\Big],
\ee
where target spacetime coordinates $X^\mu$ represents worldsheet scalars, and $\psi^\mu$ are the two component Majorana-Weyl spinors on the worldsheet: 
 \be{}
 \psi^\mu=\begin{pmatrix}\psi^\mu_- \\ \psi^\mu_+\end{pmatrix}.
 \ee 
The matrices $\rho^a$ serve as the gamma matrices on the worldsheet and obey the Clifford algebra in two dimensions. 
\be{}
\label{Cliff}
\{\rho^a,\rho^b\}=-2\eta^{ab}. 
\ee
Majorana representation of $\rho^a$ matrices is given by :
\be{rhot}
\rho^0=\begin{pmatrix}0 & -i \\ i & 0\end{pmatrix}, \quad\rho^1=\begin{pmatrix}0 & i \\ i & 0\end{pmatrix}.\ee
By varying the action with respect to the scalar and fermions on the worldsheet we get, 
\begin{eqnarray}
\label{actionvariation}
\delta S = -\frac{T}{2}\int d^2\sigma \Big[ \partial_{\a} (\delta X^{\mu} \partial^{\a} X_{\mu}) - \delta X^{\mu} \partial_a\partial^a X_{\mu}  +i\partial_a(\bar{\psi}^\mu\rho^a)\delta \psi_\mu-i\partial_a(\bar{\psi}^\mu\rho^a\delta \psi_\mu)\Big].
\end{eqnarray}
It is evident from the variation of the action that the equations of motion for scalar and fermions are given as, 
\begin{eqnarray}
\partial_a\partial^a X^{\mu} =0, ~~~ \partial_a(\bar{\psi}^\mu\rho^a)=0
\end{eqnarray}
We shall elaborate upon the boundary term shortly. 
The action \refb{taction} exhibits invariance under local diffeomorphism transformations (with parameter $\xi$) and local supersymmetry transformations (with parameter $\epsilon$).
\begin{subequations}\label{susy}
\bea{}
\delta_\xi X&=&\xi^a\p_a X, \quad  \delta_\xi\psi_\pm = \xi^a\p_a \psi_\pm+\frac{1}{2}\p_\pm\xi^\pm\psi_\pm, \\
\delta_\e X&=&\bar{\e}\psi, \quad \delta_\e\psi = -i\rho^a\p_aX\e. 
\label{sitabhog}
\eea 
\end{subequations}
Using (\ref{susy}) in the variation of the action \eqref{actionvariation} and requiring $\delta_\xi S = 0$ and $\delta_\epsilon S = 0$ separately yield:
\begin{align}\label{sup}
    \xi^\pm=\xi^\pm(\s^\pm),~~~\e^\pm=\e^\pm(\s^\pm).
\end{align}

\subsection{Boundary conditions}\label{sec2.1}
Now we return to the fermionic boundary terms explicitly given in the last term of (\ref{actionvariation})
\begin{eqnarray}\label{panipuri}
&\int d^2\sigma\, \partial_a\,(\bar{\psi}\,\rho^a\,\delta\psi)=\oint_c dl\, n_a\,\bar{\psi}\,\rho^a\,\delta\psi 
\end{eqnarray}
 In order to show that the above term vanish, without the loss of generality, we may choose the integration contour in the counterclockwise direction as $c = c_1 \oplus c_2 \oplus c_3 \oplus c_4$ where $c_1, c_2, c_3$ and $c_4$ correspond to lines defined by $\tau = +\infty$, $\sigma = 0$, $\tau = -\infty $ and $\sigma = \pi$ respectively. Moreover, we identify $n_a$ as the two dimensional normal vector to the boundaries defined by $c_1, c_2, c_3$ and $c_4$ in the following way, 
\begin{eqnarray}
n_a (c_1) = (1,0),~~ n_a (c_2) = (0,-1),~~ n_a(c_3) = (-1,0),~~ n_a(c_4) = (0,1)
\end{eqnarray}
\begin{equation}\label{boundary}
\begin{split}
\oint_c dl\, n_a\,\bar{\psi}\,\rho^a\,\delta\psi
=\int_{c_1} d\sigma\, \bar{\psi}\,\rho^0\,\delta\psi \Big|_{\tau=\infty}& -\int_{c_2} d\tau\, \bar{\psi}\,\rho^1\,\delta\psi \Big|_{\sigma=0}\\ -&\int_{c_3} d\sigma\, \bar{\psi}\,\rho^0\,\delta\psi \Big|_{\tau=-\infty}+\int_{c_4} d\tau\, \bar{\psi}\,\rho^1\,\delta\psi \Big|_{\sigma=\pi}
\end{split}
\end{equation}
Since we assume that $\psi(\tau = \pm \infty, \sigma) = 0$, the first and the third term in the right-hand side of equation \eqref{boundary} drop out. A further simplification with the second and fourth terms leads to 
\begin{align}
\oint_c dl\, n_a\,\bar{\psi}\,\rho^a\,\delta\psi
= \int d\tau\,\Big(\psi_+(\sigma)\,\delta\psi_+(\s)-\psi_-(\s)\,\delta \psi_-(\s)\Big)\Big|_{\sigma=0,\pi}=0
\end{align}
To make the above purely fermionic boundary terms vanish, we have the following two distinct conditions.
\begin{subequations}
\begin{align}\label{rasagolla}
    \psi_+(0)=\,\psi_-(0)\,, ~~~~~ &\psi_+(\pi)=\eta\,\psi_-(\pi)\,,\\ \label{rajbhog}
    \psi_+(0)=-\,\psi_-(0)\,, ~~~~~& \psi_+(\pi)=-\eta\,\psi_-(\pi)\,,
    \end{align}
\end{subequations}
where $\eta = \pm 1$. Here \eqref{rasagolla} and \eqref{rajbhog} represent $NN$ and $DD$ boundary conditions respectively, where $N$ stands for Neumann and $D$ means Dirichlet.
In order to show that \eqref{rajbhog} is indeed the Dirichlet boundary conditions for the tensile open supersymmetric strings, 
 we employ the same boundary conditions in the bosonic sector, $\delta X^{\mu} = 0$ at $\sigma = 0,\pi$ in the first equation of supersymmetry transformations \eqref{sitabhog}. 
 \begin{equation}\label{malpua}
\delta_\e X\Big|_{\s=0,\pi}=0 \qquad \implies \epsilon^+\,\psi_-=-\epsilon^-\,\psi_+.
\end{equation}
The Dirichlet boundary condition breaks half of the worldsheet supersymmetry. We ensure this fact in two different ways by imposing appropriate conditions on the supersymmetric parameters. One way of imposing such a condition is the following,
\begin{equation} \label{ns}
\begin{split}
    \epsilon^+(0)\,= \epsilon^-(0)\,, \qquad 
    \epsilon^+(\pi)= -\epsilon^-(\pi)\,.
    \end{split}
\end{equation}
Once we combine \eqref{ns} with \eqref{malpua}, we recover the Neveu-Schwarz sector of the fermionic field satisfying the Dirichlet boundary condition,
\begin{equation}\label{khirkadam1}
    \psi_+(0)=-\,\psi_-(0),
   \qquad \psi_+(\pi)=\psi_-(\pi).
\end{equation}
The other way of imposing an appropriate condition on the supersymmetric parameter is the following, 
\begin{equation}\label{ramond}
\begin{split}
    \epsilon^+(0)\,= \epsilon^-(0)\,, \qquad 
    \epsilon^+(\pi)= \epsilon^-(\pi)\,
    \end{split}
\end{equation}
This yields the Ramond sector of the fermionic field satisfying the  Dirichlet boundary conditions.
\begin{equation}\label{khirkadam2}
    \psi_+(0)=-\psi_-(0),
    \qquad  \psi_+(\pi)=-\psi_-(\pi).
    \end{equation}


\subsection{Superspace Formulation}
RNS superstring theory described by action \eqref{taction} can also be formulated in an $\mathcal{N}=(1,1)$ superspace. The superspace corresponding to the closed string worldsheet will be characterised by ($\sigma^{\pm},\theta,\Bar{\theta}$), where $\theta$ and $\Bar{\theta}$ are two Grassmannian coordinates. A general superfield in this space can be constructed as
\begin{align}
    Y(\sigma^{\pm},\theta,\Bar{\theta})=X(\sigma^{\pm})+i\theta\psi_{+}(\sigma^{\pm})+i\Bar{\theta}\psi_{-}(\sigma^{\pm})+\frac{1}{2}\theta\Bar{\theta}B(\sigma^{\pm}).
\end{align}
The auxiliary field $B$ can be set to zero without affecting the theory, as long as we are working with on-shell supersymmetry. In that case, the gauge fixed action of RNS superstring theory given in \eqref{taction} can be rewritten in terms of integral in this superspace
\begin{align}\label{superaction}
    S=-\frac{T}{2}\int~d^2\theta d^2\sigma \overline{D}YDY.
\end{align}

Now supersymmetry transformations combined with the diffeomorphism in \eqref{susy} leads us to the following transformation in the superspace coordinates
\begin{subequations}\label{sspace}
\bea{}
\delta\sigma^+&=&\xi^++i\e^+\theta, \ \ \ \delta\theta=\e^++\frac{1}{2}\theta\p_+\xi^+,\\
\delta\sigma^-&=&\xi^-+i\e^-\bar{\theta}, \ \ \ \delta\bar{\theta}=\e^-+\frac{1}{2}\bar{\theta}\p_-\xi^-.
\eea
\end{subequations}
In order to find the generators of the supersymmetry transformation in the superspace, we consider the following variation of $Y$
\begin{align}\label{sfield}
\delta Y&=(\delta\sigma^+\p_++\delta\sigma^-\p_-+\delta\theta\p_\theta+\delta\bar{\theta}\p_{\bar{\theta}})Y \nonumber\\
&=\Big[\Big(\xi^+(\s^+)\p_++\frac{1}{2}\p_+\xi^+(\s^+)\theta\p_\theta\Big)+\e^+(\s^+)\Big(\p_{\theta}+i\theta\p_+\Big)\nonumber\\
& +\Big(\xi^-(\s^-)\p_-+\frac{1}{2}\p_-\xi^-(\s^-)\bar{\theta}\p_{\bar{\theta}}\Big)+\e^-(\s^-)\Big(\p_{\bar{\theta}}+i\bar{\theta}\p_-\Big)\Big]Y \nonumber \\
&=[\mathcal{L}^+(\xi^+)+\mathcal{Q}(\e^+)+\mathcal{L}^-(\xi^-)+\bar{\mathcal{Q}}(\e^-)]Y.
\end{align}
In the above, $\mathcal{L}^+$'s give bosonic generators and $\mathcal{Q}$'s give fermionic generators. The functional forms of $\xi^{\pm}$ and $\e^{\pm}$ as given in \eqref{sup} can be expressed as Fourier expansion
\be{}\label{Fourier1}
\xi^{\pm}(\s^\pm) = \sum_{n} a_n^{\pm} e^{in\s^\pm},~~~\e^{\pm}(\s^\pm) = \sum_{r\in\mathbb{Z}+\frac{1}{2}} b_r^{\pm} e^{ir\s^\pm}.
\ee
Using \eqref{Fourier1}, the generators $\mathcal{L}^+$, $\mathcal{Q}$, $\mathcal{L}^-$ and $\bar{\mathcal{Q}}$ too can be Fourier expanded
\begin{subequations}
\bea{2}
\label{virvectorf}
\mathcal{L}^+(\xi^+)&=&-i\sum_n a^+_n \mathcal{L}^+_n,~~~\mathcal{Q}(\e^+)=\sum_{r\in\mathbb{Z}+\frac{1}{2}} b^+_r \mathcal{Q}_r,\\
\mathcal{L}^-(\xi^-)&=&-i\sum_n a^-_n \mathcal{L}^-_n,~~~\bar{\mathcal{Q}}(\e^-)=\sum_{r\in\mathbb{Z}+\frac{1}{2}} b^-_r \bar{\mathcal{Q}}_r.
\eea
\end{subequations}
From the above, we obtain the following form of the symmetry generators:
\begin{subequations}\label{shankhadeepda}
    \begin{align}
        &\mathcal{L}^+_{n}=ie^{in\sigma^+}\Big(\partial_++\frac{in}{2}\theta\partial_\theta\Big)~~~~\mathcal{Q}_r=e^{ir\sigma^+}(\partial_{\theta}+i\theta\partial_+)\\
        &\mathcal{{L}}^-_{n}=ie^{in\sigma^-}\Big(\partial_-+\frac{in}{2}\Bar{\theta}\partial_{\Bar{\theta}}\Big)~~~~\mathcal{\Bar{Q}}_r=e^{ir\sigma^-}(\partial_{\Bar{\theta}}+i\Bar{\theta}\partial_-).
    \end{align}
\end{subequations}
In the above $\partial_{\pm}=\frac{1}{2}(\partial_\tau\pm\partial_\sigma)$. They follow two copies of the Super-Virasoro algebra given by the following
\begin{subequations}
\label{svalgebra}
\begin{align}
[\mathcal{L}^+_{n},\mathcal{L}^+_{m}]=(n-m)\mathcal{L}^+_{n+m},~~~[\mathcal{L}^+_{n},\mathcal{Q}_{r}]=\Big(\frac{n}{2}-r\Big)\mathcal{Q}_{n+r},~~~\{\mathcal{Q}_{r},\mathcal{Q}_{s}\}=2\mathcal{L}^+_{r+s}.
 \\ 
[\mathcal{L}^-_{n},\mathcal{L}^-_{m}] =(n-m)\mathcal{L}^-_{n+m},~~~[\mathcal{L}^-_{n},\bar{\mathcal{Q}}_{r}]=\Big(\frac{n}{2}-r\Big)\bar{\mathcal{Q}}_{n+r},~~~\{\bar{\mathcal{Q}}_{r},\bar{\mathcal{Q}}_{s}\}=2\mathcal{L}^-_{r+s}.
\end{align}
 \end{subequations}


\subsection{Boundary in superspace coordinates}\label{dinosaur}
In this section, we recall the superspace formalism for the tensile open superstring. In bosonic open string theory, the location of the worldsheet boundary is referred to $\sigma = 0$ and $\sigma = \pi$, $\forall \tau$. When we extend the description of the worldsheet to superspace coordinates by adding Grassmann variables $\theta$ and $\bar{\theta}$, the definition of the boundary must also be extended to the entire superspace, where the boundary location must be specified in terms of Grassmann coordinates too. 
Since this work focuses exclusively on Dirichlet boundary conditions, the following discussion will be restricted to that case. Let us express the variation of the superfield in terms of the variation of bosonic and fermionic fields
\begin{equation}\label{superdirichlet}
    \delta Y(\sigma^{\pm},\theta,\Bar{\theta})=\delta X(\sigma^{\pm})+i\delta\theta\psi_{+}(\sigma^{\pm})+i\theta\delta\psi_{+}(\sigma^{\pm})+i\delta\Bar{\theta}\psi_{-}(\sigma^{\pm})+i\Bar{\theta}\delta\psi_{-}(\sigma^{\pm})
\end{equation}
Here we extend the Dirichlet condition imposed on the bosonic part $\delta X(\sigma^{\pm})|_{\sigma=0,\pi} = 0$ 
to the superfield by demanding
\begin{equation}\label{BD}
      \delta Y(\sigma^{\pm},\theta,\Bar{\theta})|_{bdy} = \delta\theta\psi_{+}(\sigma^{\pm})|_{bdy}+\theta\delta\psi_{+}(\sigma^{\pm})|_{bdy}+\delta\Bar{\theta}\psi_{-}(\sigma^{\pm})|_{bdy}+\Bar{\theta}\delta\psi_{-}(\sigma^{\pm})|_{bdy}=0
\end{equation}
The above equation \eqref{BD} tells us how the generalization of Dirichlet boundary condition to superfields should impose restrictions on the fermions for a given boundary location in superspace. In the following discussion, we obtain the location of the superspace boundaries which will lead us to the Neveu-Schwarz sector and the Ramond sectors defined in \eqref{ns} and \eqref{ramond}.  
\subsection*{Neveu Schwarz sector}
We recall the 
Neveu Schwarz sector of the fermionic part of the RNS superstring previously mentioned in
\eqref{ns}
\begin{subequations}
  \begin{equation}\label{NS1}
   \psi_{+}(0,\tau)=-\psi_{-}(0,\tau)~~\implies~~\delta\psi_{+}(0,\tau)=-\delta\psi_{-}(0,\tau)
   \end{equation} 
   \begin{equation}\label{NS2}
   \psi_{+}(\pi,\tau)=\psi_{-}(\pi,\tau)~~\implies~~\delta\psi_{+}(\pi,\tau)=\delta\psi_{-}(\pi,\tau).
\end{equation}  
\end{subequations}
Applying \eqref{NS1} on \eqref{BD} 
one gets
\begin{equation}\label{S.S.}
    \delta(\theta-\Bar{\theta})\,\psi_{+}(0,\tau)+(\theta-\Bar{\theta})\,\delta\psi_{+}(0,\tau)=0.
\end{equation}
From the above, one can see that, for non-vanishing functions  $\psi_{+}(0,\tau)$ and $\delta\psi_{+}(0,\tau)$ the LHS of \eqref{S.S.} can vanish only if $\theta=\Bar{\theta}$. That means the superspace location of this boundary is $\sigma=0,~\theta=\Bar{\theta}$. Similarly, applying \eqref{NS2} on \eqref{BD} for $\sigma=\pi$ boundary, we get
\begin{equation}\label{S.M.}
 \delta(\theta+\Bar{\theta})\, \psi_{+}(\pi,\tau)+(\theta+\Bar{\theta})\,\delta\psi_{+}(\pi,\tau)=0.
\end{equation}
That means the LHS of \eqref{S.M.} can vanish only if $\theta=-\Bar{\theta}$. Hence the superspace location of this boundary will be $\sigma=\pi,~\theta=-\Bar{\theta}$.
Hence, the two boundaries of Neveu Schwarz sector are
    \begin{align}\label{BoundaryNS} 
    \sigma=0,~~\theta=\Bar{\theta}~; \quad \sigma=\pi,~~\theta=-\Bar{\theta}.
\end{align}
For the sake of convenience in future calculations we define the superspace coordinates ($\theta, \bar{\theta}$) as 
\begin{align}    \theta=\theta_1+\theta_2~~~~\Bar{\theta}=\theta_1-\theta_2.
\end{align}
So, the boundary location as expressed in terms of these new coordinates $(\theta_1,\theta_2)$ are
\begin{align}\label{dns1}
    \sigma=0,~~~~\theta_2 = 0, \quad \sigma=\pi,~~~~\theta_1=0.
\end{align}
These superspace boundaries correspond to the superstring theory for NS sector having Dirichlet boundary conditions. 
\subsection*{Ramond Sector}
For the Ramond sector, we need to use \eqref{ramond} in \eqref{BD}. For $\s=0$, we still have \eqref{NS1} and consequently \eqref{S.S.} as well. Hence, for the Ramond sector too, the superspace boundary is located at $\sigma=0,~\theta=\Bar{\theta}$ or $\theta_2=0$. However, at $\s=\pi$, \eqref{NS2} will be replaced by
\begin{align}
    \psi_{+}(\pi,\tau)=-\psi_{-}(\pi,\tau)~~\implies~~\delta\psi_{+}(\pi,\tau)=-\delta\psi_{-}(\pi,\tau).
\end{align}
As a result, \eqref{S.M.} will be replaced by
\begin{equation}\label{S.M.2}
 \delta(\theta-\Bar{\theta}) \psi_{+}(\pi,\tau)+(\theta-\Bar{\theta})\delta\psi_{+}(\pi,\tau)=0.
\end{equation}
The inevitable consequence of \eqref{S.M.2} is again $\theta=\Bar{\theta}$. Hence, the other boundary for Ramond sector is located at $\s=\pi,~\theta=\Bar{\theta}$ or $\theta_2=0$.
Hence, the location of the superspace  boundaries for the Ramond sector is summarized as
\begin{subequations}
    \begin{align}
    \sigma=0,~~~~\theta_2 = 0\\
     \sigma=\pi,~~~~\theta_2=0
\end{align}
\end{subequations}
These superspace boundaries correspond to the superstring theory for Ramond sector having Dirichlet boundary conditions.

\subsection{Boundary compatible symmetry generators}\label{Srikakkeshwarkuchkuche}
Now, we shall study in detail how the two copies of super-Virasoro symmetry are broken due to the presence of these boundaries, and the remaining symmetry generators construct one copy of the super-Virasoro algebra. We shall explicitly show that the remaining symmetry generators satisfying one copy of the super-Virasoro algebra are compatible with the superspace boundary condition.

\medskip

As discussed in the previous section, in order to segregate the symmetry generators compatible with the boundary from those incompatible with the boundary, we need to write the generators in the basis defined in \eqref{Arjunda}, where the algebra takes the form given in \eqref{svbold}.
Now, using \eqref{shankhadeepda}, let us recast the coordinate representations of these new generators in the following way,
\begin{subequations}
\begin{align}\label{different1}    \mathbb{L}_{n}&=ie^{in\sigma^+}\Big(\partial_++\frac{in}{2}\theta\partial_\theta\Big)+ie^{in\sigma^-}\Big(\partial_-+\frac{in}{2}\Bar{\theta}\partial_{\Bar{\theta}}\Big)\\      \widetilde{\mathbb{L}}_{n}&=ie^{in\sigma^+}\Big(\partial_++\frac{in}{2}\theta\partial_\theta\Big)-ie^{in\sigma^-}\Big(\partial_-+\frac{in}{2}\Bar{\theta}\partial_{\Bar{\theta}}\Big)\\ \label{different3}
      \mathbb{Q}_{r}&=e^{ir\sigma^+}(\partial_{\theta}+i\theta\partial_+)+
e^{ir\sigma^-}(\partial_{\Bar{\theta}}+i\Bar{\theta}\partial_-)\\
          \widetilde{\mathbb{Q}}_{r}&=e^{ir\sigma^+}(\partial_{\theta}+i\theta\partial_+)-
e^{ir\sigma^-}(\partial_{\Bar{\theta}}+i\Bar{\theta}\partial_-).  
        \end{align} \label{different2}
\end{subequations}
Now, recall that the Dirichlet Neveu Schwarz sector and the Ramond sector of the tensile open superstring correspond to the superspace boundary defined as $\sigma=0,~\theta_2=0$\footnote{We choose this boundary as we encounter this boundary in the previous section for both Dirichlet NS and Dirichlet Ramond sector.}.  At this boundary these generators take the following form 
\begin{subequations}\label{jurrasicpark1}
   \begin{align} &\mathbb{L}_{n}|_{\sigma=\theta_2=0}=ie^{in\tau}\Big(\partial_\tau+\frac{in}{2}\theta_1\partial_{\theta_1}\Big),~~~~\mathbb{Q}_{r}|_{\sigma=\theta_2=0}=e^{in\tau}(\partial_{\theta_1}+\theta_1\partial_\tau),
        \\
        &\widetilde{\mathbb{L}}_{n}|_{\sigma=\theta_2=0}=ie^{in\tau}\Big(\partial_\sigma+\frac{in}{2}\theta_1\partial_{\theta_2}\Big),~~~~
        \widetilde{\mathbb{Q}}_{r}|_{\sigma=\theta_2=0}=e^{in\tau}(\partial_{\theta_2}+\theta_1\partial_\sigma).
    \end{align}\label{jurrasicpark2}
\end{subequations}

It is crucial to note that 
since the generators $\widetilde{\mathbb{L}}_{n}$ and $\widetilde{\mathbb{Q}}_{r}$ 
distort the the superspace boundary defined  by $\sigma=0$ and $\theta_2=0$ (due to the presence of $\partial_\sigma$ and $\partial_{\theta_2}$ term in both of them),  hence they must be ruled out. However, the generators $\mathbb{L}_{n}$ and $\mathbb{Q}_{r}$ won't distort the boundary, and consequently those symmetries will remain both in the Dirichlet Neveu Schwarz and Dirichlet Ramond sectors. 


To determine the generators respecting the superspace Dirichlet boundary condition in Neveu Schwarz and the Ramond sectors, for which the bosonic part corresponds to $\sigma = \pi$, requires special attention\footnote{We no longer pay attention to the generators $\widetilde{\mathbb{L}}_{n}$ and $\widetilde{\mathbb{Q}}_{r}$ since they are eliminated by the boundary $\{\s=0,~\theta_2=0\}$, a boundary common to both NS and R sector.}.

\subsection*{Ramond Sector}
In the Ramond sector, the index $r$ of the generators $\mathbb{Q}_{r}$ happens to be an integer. If we locate the other boundary at $\sigma=\pi$ and $\theta_2=0$ the generators $\mathbb{L}_{n}$ and $\mathbb{Q}_{r}$ take the following form, 
\begin{subequations}\label{Ramond}
    \begin{align}\label{ramond1}
        &\mathbb{L}_{n}|_{\sigma=\pi}=ie^{in\pi}e^{in\tau}\Big(\partial_++\frac{in}{2}\theta\partial_\theta\Big)+ie^{-in\pi}e^{in\tau}\Big(\partial_-+\frac{in}{2}\Bar{\theta}\partial_{\Bar{\theta}}\Big)\\
        &~~~~\mathbb{Q}_{r}|_{\sigma=\pi}=e^{ir\pi}e^{ir\tau}(\partial_{\theta}+i\theta\partial_+)+
e^{-ir\pi}e^{ir\tau}(\partial_{\Bar{\theta}}+i\Bar{\theta}\partial_-)
    \end{align}
\end{subequations}
Now, since for Ramond sector, both $n$ and $r$ are integers, $e^{2\pi ir}=1$ i.e., $e^{ir\pi}=e^{-ir\pi}=(-1)^r$ (this is true for $n$ too, of course). Hence, in the $\sigma=\pi,\theta_2=0$ boundary, the generators $\mathbb{L}_{n}$ and $\mathbb{Q}_r$ becomes
\begin{subequations}\label{Ramond2}
\begin{align}
\mathbb{L}_{n}|_{\sigma=\pi,\theta_2=0}& = i(-1)^ne^{in\tau}\Big(\partial_\tau+\frac{in}{2}\theta_1\partial_{\theta_1}\Big) \\
\mathbb{Q}_{r}|_{\sigma=\pi,\theta_2=0}&= (-1)^re^{ir\tau}\big(\partial_{\theta_1}+i\theta_1\partial_\tau\big)
\end{align}
\end{subequations}

As we can see from \eqref{Ramond2}, none of the generators $\mathbb{L}_n$ and $\mathbb{Q}_r$ contain any $\partial_\sigma$ or $\partial_{\theta_2}$ at the boundary located at $\sigma=\pi$, $\theta_2=0$. Hence, one can say that for Ramond sector, together, the boundaries located at $\sigma=0$, $\theta_2=0$ and $\sigma=\pi$, $\theta_2=0$ in the superspace allow symmetry generators $\mathbb{L}_n$ and $\mathbb{Q}_r$ with $n,r\in\mathbb{Z}$ to persist.
\subsection*{Neveu-Schwarz Sector}
In the Neveu-Schwarz (NS) sector we encounter a different scenario. Here, $n\in\mathbb{Z}$ but $r\in\mathbb{Z}+\frac{1}{2}$. As a result, the generators $\mathbb{Q}_r$s will encounter a different scenario. Let us look into the expression of $\mathbb{Q}_{r}$ as given \eqref{Ramond}.
Now, we know that for $r\in\mathbb{Z}+\frac{1}{2}$, $e^{2\pi i r}=e^{\pi i}$. Which means $e^{\pi i r}=e^{i\pi/2}=i$ and $e^{-\pi i r}=e^{-i\pi/2}=-i$. With this, the generators $\mathbb{Q}_r$ take the following form at $\sigma=\pi$
\begin{equation}
\begin{split}
   \mathbb{Q}_{r}|_{\sigma=\pi}=ie^{ir\tau}\big(\partial_{\theta}+i\theta\partial_+- \partial_{\Bar{\theta}}-i\Bar{\theta}\partial_-\big)
\end{split}
\end{equation}
Now, here we see the difference. If we want to extend the $\sigma=\pi$ boundary to the $\theta=\bar{\theta}$ ($\theta_2=0$), then we see that at the boundary located at $\sigma=\pi$, $\theta_2=0$ the generators $\mathbb{Q}_{r}$ becomes
\begin{align}
    \mathbb{Q}_{r}|_{\sigma=\pi,\theta_2=0}=ie^{ir\tau}\big(\partial_{\theta_2}+i\theta_1\partial_\sigma\big).
\end{align}
As we can see that in this boundary $\mathbb{Q}_{r}$ acquires both $\partial_\sigma$ as well as $\partial
_{\theta_2}$ which makes this symmetry generator incompatible with the $\sigma=\pi,\theta_2=0$ boundary.\par
\medskip

Now, let us try a different location for the $\sigma=\pi$ boundary in the superspace i.e., $\theta=-\bar{\theta}$ (where $\theta_1=0$) instead of $\theta=\bar{\theta}$. Then, at this new location we see the following 
\begin{align}
    \mathbb{Q}_{r}|_{\sigma=\pi,\theta_1=0}=ie^{ir\tau}\big(\partial_{\theta_2}+i\theta_2\partial_\tau\big).
\end{align}
Here we see that the generator $\mathbb{Q}_{r}$ doesn't contain any $\partial_\sigma$ or $\partial_{\theta_1}$, implying that this boundary is left untouched by this symmetry generators.
\medskip

Before we jump into conclusions, we should ask ourselves one question: are the $\mathbb{L}_n$'s compatible with this newly defined boundary? In order to see this we have to simply put $\theta_1=0$ instead of $\theta_2=0$ in the expression of $\mathbb{L}_n|_{\sigma=\pi}$ as given in \eqref{ramond1}. This leads us to 
\begin{align}
    \mathbb{L}_{n}|_{\sigma=\pi,\theta_1=0}=
i(-1)^ne^{in\tau}\Big(\partial_\tau+\frac{in}{2}\theta_2\partial_{\theta_2}\Big).
\end{align}
Hence, $\mathbb{L}_{n}$ generators too, like $\mathbb{Q}_r$'s, free from any $\partial_\sigma$ or $\partial_{\theta_1}$ terms at this boundary, and hence perfectly compatible with this boundary. Hence we see that for NS sector the boundaries \{$\sigma=0$, $\theta_1=0$\} and \{$\sigma=\pi$, $\theta_2=0$\} allow the symmetry generators $\mathbb{L}_{n}$ and $\mathbb{Q}_r$ with $n\in\mathbb{Z}$ and $r\in\mathbb{Z}
+\frac{1}{2}$ to persist. 
\medskip

Finally, from \eqref{svbold}, we observe that the allowed symmetry generators in the RNS sectors, consistent with both boundaries, form a single copy of the super-Virasoro algebra without the central term, as given in \eqref{bsv}, where it is derived from the field-theoretic perspective.
\section{Different Sets of Boundaries in Superspace} \label{different set if boundaries in superspace}
In section \ref{BasantaBiswas}, we saw how super-Virasoro algebra can be rewritten in a different basis \eqref{svbold}. From \eqref{svbold} we also saw that $\mathbb{Q}~ \text{and}~ \tilde{\mathbb{Q}}$ separately gives one copy of super-Virasoro algebra with $\mathbb{L}$. However, $\{\mathbb{Q},\tilde{\mathbb{Q}}\}$ gives rise to $\tilde{\mathbb{L}}$ not present in the single copy of Virasoro algebra with $\mathbb{L}$
. The implication was that, once we introduce the boundary, either one of $\mathbb{Q}$ or $\widetilde{\mathbb{Q}}$ would survive but not both. While discussing the SCCA$_H$, we saw something very similar too; while writing in a different basis it became clear that once we introduce boundary in the superspace, either $\mathcal{H}$ or $\mathcal{I}$ survives but not both. While studying the superspaces for tensile superstrings in \ref{Srikakkeshwarkuchkuche} and homogeneous tensionless superstrings in \ref{hijibijbij2}  we also found the followings:
\begin{enumerate}
    \item  For Neveau Schwarz sector in tensile superstrings, the boundaries \{$\sigma = 0, \theta_2 = 0$\} and \{$\sigma = \pi, \theta_1=0$\} allow the symmetry generators $\mathbb{L}_n$s and $\mathbb{Q}_r$s where $r\in\mathbb{Z}+\frac{1}{2}$.
    \item For Ramond sector in tensile superstrings, the boundaries \{$\sigma = 0, \theta_2 = 0$\} and \{$\sigma = \pi, \theta_2=0$\} allow the symmetry generators $\mathbb{L}_n$s and $\mathbb{Q}_r$s where $r\in\mathbb{Z}$.
    \item For Neveau Schwarz sector in homogeneous tensionless superstrings, the boundaries \{$\sigma = 0, \theta_2 = 0$\} and \{$\sigma = \pi, \theta_1=0$\} allow the symmetry generators $\mathcal{O}_n$s and $\mathcal{P}_r$s and $\mathcal{H}_r$s where $r\in\mathbb{Z}+\frac{1}{2}$.
\end{enumerate}
In this appendix we explore other possible sets of  boundaries introduced in the superspace and their impact on the symmetry generators. As we will see, this will provide us with a more complete picture of the effect of the boundaries on the symmetries. 

\subsection{Super-Virasoro Algebra}

Let us introduce a boundary at $\sigma=0 ~\text{and} ~ \theta_1=0~ (\text{i.e.}~ \theta=-\bar{\theta})$ in the superspace. It is easy to see that for this boundary, the generators $\mathbb{Q}$'s are eliminated and $\widetilde{\mathbb{Q}}$ survive, (just in the similar way at $\sigma=0 ~\text{and} ~ \theta_2=0~ (\text{i.e.}~ \theta=\bar{\theta})$, generators $\widetilde{\mathbb{Q}}$'s are eliminated and $\mathbb{Q}$ survives) as demonstrated below
  \begin{align}
        \mathbb{Q}_r|_{\sigma=\theta_1=0}=e^{ir\tau}(\partial_{\theta_1}+i\,\theta_2\partial_\sigma),~~~
        \widetilde{\mathbb{Q}}_r|_{\sigma= \theta_1=0}=e^{ir\tau}(\partial_{\theta_2}+i\,\theta_2\partial_\tau).
        \end{align}
Since $\mathbb{Q}_r|_{\sigma=\theta_1=0}$ contain $\partial_\sigma$ and $\partial_{\theta_1}$, $\mathbb{Q}_r$s do not survive, while $\widetilde{\mathbb{Q}}_r$, being free from $\partial_\sigma$ and $\partial_{\theta_1}$ at $\sigma=\theta_1=0$, survive. For the same reason, in this boundaries, $\mathbb{L}_n$s survives but $\widetilde{\mathbb{L}}_n$s do not
\begin{align} \mathbb{L}_{n}|_{\sigma=\theta_1=0}&=ie^{in\tau}\Big(\partial_\tau+\frac{in}{2}\theta_2\partial_{\theta_2}\Big),~~~\widetilde{\mathbb{L}}_{n}|_{\sigma=\theta_1=0}=ie^{in\tau}\Big(\partial_\sigma+\frac{in}{2}\theta_2\partial_{\theta_1}\Big).
    \end{align}
Here too, one encounters the Ramond Sectors and Neveu Schwarz sectors. 
\subsection*{$\widetilde{\mathbb{Q}}_r$ for Ramond Sector}
For Ramond sector, where $r\in\mathbb{Z}$, the other boundary should be $\s=\pi$ and $\theta_1=0$. In this boundary, one can see that the $\widetilde{\mathbb{Q}}_r$ and $\mathbb{L}_n$s are free from $\partial_\sigma$ and $\partial_{\theta_1}$
\begin{align}
    \mathbb{L}_n|_{\s=\pi,\theta_1=0}=i(-1)^ne^{in\tau}\Big(\partial_\tau+\frac{in}{2}\theta_2\partial_{\theta_2}\Big),~~~\widetilde{\mathbb{Q}}_r|_{\s=\pi,\theta_1=0}=(-1)^{r}e^{ir\t}(\partial_{\theta_2}+i\,\theta_2\partial_\t).
\end{align}
Hence, together the two boundaries $\sigma=0, \theta_1=0$ and $\s=\pi, \theta_1=0$ allow super-Virasroro generators $\mathbb{L}_n$ and $\widetilde{\mathbb{Q}}_r$ with $n,r\in\mathbb{Z}$ (Ramond sector) to survive in superspace.
\subsection*{$\widetilde{\mathbb{Q}}_r$ for Neveu Schwarz Sector}
For Neveu Schwarz sector, where $r\in\mathbb{Z}+\frac{1}{2}$, the other boundary should be $\s=\pi$ and $\theta_2=0$. In this boundary, one can see that the $\widetilde{\mathbb{Q}}_r$ and $\mathbb{L}_n$s are free from $\partial_\sigma$ and $\partial_{\theta_2}$
\begin{align}
    \mathbb{L}_n|_{\s=\pi,\theta_1=0}=i(-1)^ne^{in\tau}\Big(\partial_\tau+\frac{in}{2}\theta_1\partial_{\theta_1}\Big),~~~\widetilde{\mathbb{Q}}_r|_{\s=\pi,\theta_1=0}=ie^{ir\t}(\partial_{\theta_1}+i\,\theta_1\partial_\t).
\end{align}
Hence, together the two boundaries $\sigma=0, \theta_1=0$ and $\s=\pi, \theta_1=0$ allow super-Virasroro generators $\mathbb{L}_n$ and $\widetilde{\mathbb{Q}}_r$ with $n\in\mathbb{Z}$ and $r\in\mathbb{Z}+\frac{1}{2}$ (Neveu Schwarz sector) to survive in superspace.
\subsection{BSCCA$_H$}
In the section \ref{hijibijbij2} we discussed about the symmetries on the Homogeneous Carrollian superspace where boundaries are at \{$\sigma = 0, \theta_2 = 0$\} and \{$\sigma = \pi, \theta_1=0$\}.
The compatible symmetry generators were $\mathcal{O}_n$, $\mathcal{P}_n$ and $\mathcal{H}_r$, and they belonged to the Neveu Schwarz Sector. Here we explore other boundaries. Here we discuss three other possible choices of boundaries which will lead us other possibilities
\subsection*{$\mathcal{H}_r$ for Ramond Sector}
Let us consider the boundaries \{$\sigma = 0, \theta_2 = 0$\} and \{$\sigma = \pi, \theta_2=0$\}. It can be straightforward to show that the generators $\mathcal{O}_n$, $\mathcal{P}_n$ and $\mathcal{H}_r$ as given in \eqref{BSBMS} are free from any $\partial_\sigma$ and $\partial_{\theta_2}$ terms at those boundaries as long as $r\in\mathbb{Z}$. One can also check that $\mathcal{I}_r$s as given in \eqref{T-Rex} does contain $\partial_{\theta_2}$ term  at both the boundaries for $r\in\mathbb{Z}$, making them incompatible at those boundaries.
Hence the symmetry generators for Ramond sector ($r\in\mathbb{Z}$) in these boundaries remain same.
\subsection*{$\mathcal{I}_r$ for Neveu Schwarz and Ramond Sectors}
Let us consider the boundary \{$\sigma=0,~\theta_1=0$\}. Looking at the generators $\mathcal{O}_n$, $\mathcal{P}_n$ and $\mathcal{I}_r$ as given in \eqref{pteranodon} and \eqref{T-Rex} it can be immediately seen that $\mathcal{O}_n$, $\mathcal{P}_n$, and $\mathcal{I}_r$ are free of $\partial_\sigma$ and $\partial_{\theta_1}$ terms at \{$\sigma=0,~\theta_1=0$\} boundary which means this boundary allow $\mathcal{O}_n$, $\mathcal{P}_n$, and $\mathcal{I}_r$ symmetries. However, $\mathcal{H}_r$s as given in \eqref{T-Rex} have $\partial_{\theta_1}$ at \{$\sigma=0,~\theta_1=0$\}, making them incompatible with this boundary. Now, if we set the other boundary at \{$\sigma=\pi,~\theta_2=0$\}, one can see that $\mathcal{O}_n$, $\mathcal{P}_n$ and $\mathcal{I}_r$ are free of the $\partial_\sigma$ and $\partial_{\theta_2}$
terms provided $r\in\mathbb{Z}+\frac{1}{2}$, i.e. the generators belong to the Neveu Schwarz sector. That means together the superspace boundaries \{$\sigma=0,~\theta_1=0$\} and \{$\sigma=\pi,~\theta_2=0$\} allow $\mathcal{O}_n$, $\mathcal{P}_n$ and $\mathcal{I}_r$ with $r\in\mathbb{Z}+\frac{1}{2}$ as symmetry generators.
\medskip

However, once we set the other boundary at \{$\sigma=\pi,~\theta_1=0$\}, we see that $\mathcal{O}_n$, $\mathcal{P}_n$ and $\mathcal{I}_r$ are free of any $\partial_\sigma$ and $\partial_{\theta_1}$  term at the other boundary
terms provided $r\in\mathbb{Z}$ i.e. generators belong to Ramond sector now. Hence, together the superspace boundaries \{$\sigma=0,~\theta_1=0$\} and \{$\sigma=\pi,~\theta_1=0$\} allow $\mathcal{O}_n$, $\mathcal{P}_n$ and $\mathcal{I}_r$ with $r\in\mathbb{Z}$ as symmetry generators. 
\subsection*{Limit from Super-Virasoro}
It is important to note that the superspace boundaries which permits $\mathbb{Q}_r$ ($\widetilde{\mathbb{Q}}_r$) are identical to the boundaries allowing $\mathcal{H}_r$ ($\mathcal{I}_r$) generators. 
We previously saw that $\mathcal{H}$ can be derived from $\mathbb{Q}$ as $\mathcal{H}_r=\sqrt{\epsilon}\mathbb{Q}_r$. Similarly, one can also see that $\mathcal{I}_r$ can be derived from $\widetilde{\mathbb{Q}}_r$ as  $\mathcal{I}_r=\sqrt{\epsilon}\widetilde{\mathbb{Q}}_r$. Hence these separate set of boundaries results into different sets of generators which for relativistic case satisfy one copy of super-Virasoro algebra, and for the homogeneous Carrollian superspace satisfy BSCCA. In our study we chose to work with $\theta_2=0$ boundary because it was compatible with the Dirichlet boundary condition as shown in section \eqref{dinosaur}.
\newpage
\bibliographystyle{JHEP}
\bibliography{ccft}
\end{document}